\documentclass{article}
\begin{document}
{\LARGE
\begin{center}
{\bf
Excited ${\bf (70,1^-)}$ baryon resonances in the relativistic quark
model}
\end{center}
}
\vskip3ex
\noindent
S.M. Gerasyuta and E.E. Matskevich

\vskip2ex
\noindent
Department of Theoretical Physics, St. Petersburg State University,
198904, St. Petersburg, Russia

\noindent
Department of Physics, LTA, 194021, St. Petersburg, Russia

\vskip4ex
\begin{center}
{\bf Abstract}
\end{center}
\vskip4ex
{\large
The relativistic three-quark equations of the ${\bf (70,1^-)}$ baryons
are found in the framework of the dispersion relation technique.
The approximate solutions of these equations using the method based
on the extraction of leading singularities of the amplitude are
obtained. The calculated mass values of the ${\bf (70,1^-)}$ multiplet
are in good agreement with the experimental ones.
\vskip2ex
\noindent
e-mail address: gerasyuta@SG6488.spb.edu

\noindent
e-mail address: matskev@pobox.spbu.ru
\vskip2ex
{\bf 1. Introduction.}
\vskip 5mm
At low energies, typical for baryon spectroscopy, QCD does not admit
a perturbative expansion in the strong coupling constant. In 1974
't Hooft [1] suggested a perturbative expansion of QCD in terms of the
parameter $1/N_c$ where $N_c$ is the number of colors. This suggestion
together with the power counting rules of Witten [2] has lead to the
$1/N_c$ expansion method which allows to systematically analyse baryon
properties. The success of the method stems from the discovery that
the ground state baryons have an exact contracted $SU(2N_f)$ symmetry
when $N_c \to \infty$ [3, 4], $N_f$ being the number of flavors. For
$N_c \to \infty$ the baryon masses are degenerated. For large $N_c$ the
mass splitting starts at order $1/N_c$. Operator reduction rules simplify
the $1/N_c$ expansion [5, 6].

A considerable amount of work has been devoted to the ground state
baryons, described by the ${\bf 56}$ representation of $SU(6)$
[7-11].

The excited states belonging to ${\bf (56,L)}$ multiplets are rather
simple and can be studied by analogy with the ground state. In this
case both the orbital and the spin-flavor parts of the wave functions
are symmetric. Explicit forms for such wave functions were given in
Ref [12] for the ${\bf (56,4^+)}$ multiplet.

Together with color part, they generate antisymmetric wave functions.
Naturally, it turned out that the splitting starts at order $1/N_c$
as for the ground state.

The states belonging to ${\bf (70,L)}$ multiplets are apparently more
difficult. In this case the general practice was to split the baryon
into excited quark and a symmetric core, the latter being either in
the ground state for the $N=1$ band or in an excited state for the
$N\ge 2$ bands. Recently Matagne and Stancu have suggested the new
approach [13] for the excited ${\bf (70,1^-)}$ multiplet. They solved
the problem by removing the splitting of generators and using
orbital-spin-flavor wave functions.

The excited baryons are considered as bound states. The basic conclusion
is that the first order correction to the baryon masses is order $1/N_c$
instead of order $N_c^0$, as previously found. The conceptual difference
between the ground state and the excited states is therefore removed.

In the series of papers [14-18] a practical treatment of relativistic
three-hadron systems have been developed. The physics of three-hadron
system is usefully described in term of the pairwise interactions
among the three particles. The theory is based on the two principles
of unitarity and analyticity, as applied to the two-body subenergy
channels. The linear integral equations in a single variable are obtained
for the isobar amplitudes. Instead of the quadrature methods of obtaining
solution the set of suitable functions is identified and used as basis
set for the expansion of the desired solutions. By this means the couple
integral equation are solved in terms of simple algebra.

In our papers [19, 20] relativistic generalization of the three-body
Faddeev equations was obtained in the form of dispersion relations in the
pair energy of two interacting particles. The mass spectrum of $S$-wave
baryons including $u$, $d$, $s$-quarks was calculated by a method based
on isolating the leading singularities in the amplitude. We searched for
the approximate solution of integral three-quark equations by taking
into account two-particle and triangle singularities, all the weaker ones
being neglected. If we considered such an approximation, which corresponds
to taking into account two-body and triangle singularities, and defined
all the smooth functions of the subenergy variables (as compared with the
singular part of the amplitude) in the middle point of the physical
region of Dalitz-plot, then the problem was reduced to the one of solving
a system of simple algebraic equations.

Section 2 is devoted to the construction of the orbital-spin-flavor
wave functions for the ${\bf (70,1^-)}$ multiplet.

In Section 3 the relativistic three-quark equations are constructed in
the form of the dispersion relation over the two-body subenergy.

In Section 4 the systems of equations for the reduced amplitudes are
derived.

Section 5 is devoted to the calculation results for the $P$-wave baryons
mass spectrum (Tables I-VI).

In Conclusion, the status of the considered model is discussed.

In Appendix A the wave functions of $P$-wave baryons are given.

In Appendix B the relativistic three-particle integral equations for the
${\bf (70,1^-)}$ multiplet are constructed.

In Appendix C the reduced equations for the ${\bf (70,1^-)}$ multiplet are
obtained.

\vskip 5mm
{\bf 2. The wave function of ${\bf (70,1^-)}$ excited states.}
\vskip 5mm
Here we deal with a three-quark system having one unit of orbital
excitation. Then the orbital part of wave function must have a mixed
symmetry. The spin-flavor part of wave function must have the same
symmetry in order to obtain a totally symmetric state in the
orbital-spin-flavor space.

For the sake of simplicity we derived the wave functions for the
$(10,2)$ decuplets. The fully symmetric wave function for the decuplet
state can be constructed as [21].

$$\varphi=\frac{1}{\sqrt{2}}\left(
\varphi_{MA}^{SU(6)}\varphi_{MA}^{O(3)}+
\varphi_{MS}^{SU(6)}\varphi_{MS}^{O(3)}
\right).\eqno (1)$$

\newpage
Then we obtain:

$$\varphi=\frac{1}{\sqrt{2}}\varphi_{S}^{SU(3)}\left(
\varphi_{MA}^{SU(2)}\varphi_{MA}^{O(3)}+
\varphi_{MS}^{SU(2)}\varphi_{MS}^{O(3)}
\right),\eqno (2)$$

\noindent
here $MA$ and $MS$ define the mixed antisymmetric and symmetric part
of wave function,

$$\varphi_{MA}^{SU(6)}=\varphi_{S}^{SU(3)}\varphi_{MA}^{SU(2)},\quad\quad
\varphi_{MS}^{SU(6)}=\varphi_{S}^{SU(3)}\varphi_{MS}^{SU(2)}.\eqno (3)$$

The functions $\varphi_{MA}^{SU(2)}$, $\varphi_{MS}^{SU(2)}$,
$\varphi_{MA}^{O(3)}$, $\varphi_{MS}^{O(3)}$ are following:

$$\varphi_{MA}^{SU(2)}=\frac{1}{\sqrt{2}}\left(
\uparrow \downarrow \uparrow-\downarrow \uparrow \uparrow
\right),\quad\quad
\varphi_{MS}^{SU(2)}=\frac{1}{\sqrt{6}}\left(
\uparrow \downarrow \uparrow+\downarrow \uparrow \uparrow-
2\uparrow \uparrow \downarrow
\right),\eqno (4)$$

$$\varphi_{MA}^{O(3)}=\frac{1}{\sqrt{2}}\left(
010-100
\right),\quad\quad
\varphi_{MS}^{O(3)}=\frac{1}{\sqrt{6}}\left(
010+100-2\cdot 001
\right).\eqno (5)$$

$\uparrow$ and $\downarrow$ determine the spin directions. $1$ and $0$
correspond to the excited or nonexcited quarks. The three projections
of quark orbital moment are $l_z=1, 0, -1$. The $(10,2)$ multiplet with
$J^p=\frac{3}{2} ^{-}$ can be obtained using the spin $S=\frac{1}{2}$
and $l_z=1$, but the $(10,2)$ multiplet with $J^p=\frac{1}{2} ^{-}$
is determined by the spin $S=\frac{1}{2}$ and $l_z=0$.

We use the functions (4), (5) and construct the $SU(3)$-function for
each particle of multiplet.

For instance, the $SU(3)$-function for $\Sigma^{+}$-hyperon of decuplet
have following form:

$$\varphi_{S}^{SU(3)}=\frac{1}{\sqrt{3}}\left(
usu+suu+uus
\right).\eqno (6)$$

Then we obtain the $SU(6)\times O(3)$-function for $\Sigma^{+}$ of
the $(10,2)$ multiplet:

$$\varphi_{\Sigma^{+}(10,2)}=\frac{\sqrt{6}}{18}\left(
2\{u^1\downarrow u\uparrow s\uparrow\}+
\{s^1\downarrow u\uparrow u\uparrow\}-\right.$$
$$\left.-\{u^1\uparrow u\downarrow s\uparrow\}-
\{u^1\uparrow u\uparrow s\downarrow \}-
\{s^1\uparrow u\uparrow u\downarrow \}\right).\eqno (7)$$

Here the parenthesis determine the symmetrical function:

$$\{{abc}\}\equiv abc+acb+bac+cab+bca+cba.\eqno (8)$$

The wave functions of $\Sigma^{0}$- ¨ $\Sigma^{-}$-hyperons can be
constructed by similar way.

For the $\Delta$ baryon of $(10,2)$ multiplet the wave function can be
obtained if we replace by $u\leftrightarrow s$ quarks.

For the $\Delta^{++}$-isobar the wave function is following:

$$\varphi_{\Delta^{++}(10,2)}=\frac{\sqrt{2}}{6}\left(
\{u^1\downarrow u\uparrow u\uparrow\}
-\{u^1\uparrow u\uparrow u\downarrow\}
\right).\eqno (9)$$

For the $\Xi^{0,-}$ state of the $(10,2)$ multiplet the wave function is
similar to the $\Sigma^{+,-}$ state with the replacement by
$u\leftrightarrow s$ or $d\leftrightarrow s$. The wave function for the
$\Omega^{-}$ of the $(10,2)$ decuplet is determined as the $\Delta^{++}$
state with the replacement by $u\rightarrow s$ quarks.

The wave function and the method of construction for the multiplets
$(8,2)$, $(8,4)$ and $(1,2)$ are similar (see the Appendix A).

\newpage
{\bf 3. The three-quark integral equations for the ${\bf (70,1^-)}$
multiplet.}
\vskip 5mm
By consideration of the construction of ${\bf (70,1^-)}$ multiplet integral
equations we need to using the projectors for the
different diquark states. The projectors to the symmetric and antisymmetric
states can be obtained as:

$$\frac{1}{2}\left(q_1 q_2+q_2 q_1\right),\quad\quad
\frac{1}{2}\left(q_1 q_2-q_2 q_1\right).\eqno (10)$$

The spin projectors are following:

$$\frac{1}{2}\left(\uparrow\downarrow+\downarrow\uparrow
\right),\quad\quad
\frac{1}{2}\left(\uparrow\downarrow-\downarrow\uparrow
\right).\eqno (11)$$

The orbital moment excitation projectors are:

$$\frac{1}{2}\left(10+01\right),\quad\quad
\frac{1}{2}\left(10-01\right).\eqno (12)$$

One can obtain the four types of totally symmetric projectors:

$$S=S\cdot S\cdot S=\frac{1}{8}\left(q_1 q_2+q_2 q_1\right)
\left(\uparrow\downarrow+\downarrow\uparrow\right)\left(10+01\right),
\eqno (13)$$

$$S=S\cdot A\cdot A=\frac{1}{8}\left(q_1 q_2+q_2 q_1\right)
\left(\uparrow\downarrow-\downarrow\uparrow\right)\left(10-01\right),
\eqno (14)$$

$$S=A\cdot A\cdot S=\frac{1}{8}\left(q_1 q_2-q_2 q_1\right)
\left(\uparrow\downarrow-\downarrow\uparrow\right)\left(10+01\right),
\eqno (15)$$

$$S=A\cdot S\cdot A=\frac{1}{8}\left(q_1 q_2-q_2 q_1\right)
\left(\uparrow\downarrow+\downarrow\uparrow\right)\left(10-01\right).
\eqno (16)$$

We use these projectors for the consideration of various diquarks:

\vskip 4mm
\noindent
$u^1\uparrow s\downarrow\hskip 2mm :$

$$\frac{a^{0S}_{1}}{8}
\left(u^1\uparrow s\downarrow+u^1\downarrow s\uparrow+
s^1\uparrow u\downarrow+s^1\downarrow u\uparrow+
u\uparrow s^1\downarrow+u\downarrow s^1\uparrow+
s\uparrow u^1\downarrow+s\downarrow u^1\uparrow \right)$$
$$+\frac{a^{1S}_{0}}{8}
\left(u^1\uparrow s\downarrow-u^1\downarrow s\uparrow+
s^1\uparrow u\downarrow-s^1\downarrow u\uparrow-
u\uparrow s^1\downarrow+u\downarrow s^1\uparrow-
s\uparrow u^1\downarrow+s\downarrow u^1\uparrow \right)$$
$$+\frac{a^{0S}_{0}}{8}
\left(u^1\uparrow s\downarrow-u^1\downarrow s\uparrow-
s^1\uparrow u\downarrow+s^1\downarrow u\uparrow+
u\uparrow s^1\downarrow-u\downarrow s^1\uparrow-
s\uparrow u^1\downarrow+s\downarrow u^1\uparrow \right)$$
$$+\frac{a^{1S}_{1}}{8}
\left(u^1\uparrow s\downarrow+u^1\downarrow s\uparrow-
s^1\uparrow u\downarrow-s^1\downarrow u\uparrow-
u\uparrow s^1\downarrow-u\downarrow s^1\uparrow+
s\uparrow u^1\downarrow+s\downarrow u^1\uparrow \right),
\eqno (17)$$

\noindent
$u^1\uparrow s\uparrow\hskip 2mm :$

$$\frac{a^{0S}_{1}}{4}
\left(u^1\uparrow s\uparrow+s^1\uparrow u\uparrow+
u\uparrow s^1\uparrow+s\uparrow u^1\uparrow\right)+$$

$$+\frac{a^{1S}_{1}}{4}
\left(u^1\uparrow s\uparrow-s^1\uparrow u\uparrow-
u\uparrow s^1\uparrow+s\uparrow u^1\uparrow
\right),\eqno (18)$$

\noindent
$u\uparrow s\downarrow\hskip 2mm :$

$$\frac{a^{0S}_{1}}{4}
\left(u\uparrow s\downarrow+u\downarrow s\uparrow+
s\uparrow u\downarrow+s\downarrow u\uparrow
\right)+$$

$$+\frac{a^{0S}_{0}}{4}
\left(u\uparrow s\downarrow-u\downarrow s\uparrow-
s\uparrow u\downarrow+s\downarrow u\uparrow
\right),\eqno (19)$$

\noindent
$u\uparrow s\uparrow\hskip 2mm :$

$$\frac{a^{0S}_{1}}{2}
\left(u\uparrow s\uparrow+s\uparrow u\uparrow
\right).\eqno (20)$$

Here the down index determines the value of spin projection,
and the up index corresponds to the value of orbital moment.

If we consider the flavors $u$, $d$ that the results (21)-(24) are similar
to (17)-(20) with the replacement by $s\rightarrow d$ and the amplitudes
$a^{0}_{1}$, $a^{1}_{0}$, $a^{0}_{0}$, $a^{1}_{1}$.

\vskip 4mm
\noindent
$u^1\uparrow u\downarrow\hskip 2mm :$

$$\frac{a^{0}_{1}}{4}
\left(u^1\uparrow u\downarrow+u^1\downarrow u\uparrow+
u\uparrow u^1\downarrow+u\downarrow u^1\uparrow
\right)+$$

$$+\frac{a^{1}_{0}}{4}
\left(u^1\uparrow u\downarrow-u^1\downarrow u\uparrow-
u\uparrow u^1\downarrow+u\downarrow u^1\uparrow
\right),\eqno (21)$$

\noindent
$u^1\uparrow u\uparrow\hskip 2mm :$

$$\frac{a^{0}_{1}}{2}
\left(u^1\uparrow u\uparrow+u\uparrow u^1\uparrow
\right),\eqno (22)$$

\noindent
$u\uparrow u\downarrow\hskip 2mm :$

$$\frac{a^{0}_{1}}{2}
\left(u\uparrow u\downarrow+u\downarrow u\uparrow
\right),\eqno (23)$$

\noindent
$u\uparrow u\uparrow\hskip 2mm :$

$$a^{0}_{1}\hskip 1mm u\uparrow u\uparrow .\eqno (24)$$

Here we consider the projection of orbital moment $l_z=+1$. We use only
diquarks $1^+$, $0^+$, $2^-$, $1^-$. If we consider the $l_z=0$ and
$l_z=-1$, that we obtain the other diquarks: $1^+$, $0^+$, $1^-$, $0^-$.
In our model the five types of diquarks $1^+$, $0^+$, $2^-$, $1^-$, $0^-$
are constructed.

For the sake of simplicity we derive the relativistic Faddeev equations
using the $\Sigma$-hyperon with $J^p=\frac{3}{2} ^{-}$ of the (10,2)
multiplets. We use the graphic equations for the functions $A_J(s,s_{ik})$
[19, 20]. In order to represent the amplitude $A_J(s,s_{ik})$ in the form
of dispersion relation, it is necessary to define the amplitudes of
quark-quark interaction $b_J(s_{ik})$. The pair quarks amplitudes
$qq\rightarrow qq$ are calculated in the framework of the dispersion
$N/D$ method with the input four-fermion interaction with quantum numbers
of the gluon [21]. We use results of our relativistic quark model [22]
and write down the pair quark amplitudes in the form:

$$b_J(s_{ik})=\int\limits_{(m_i+m_k)^2}^{\infty}\hskip2mm
\frac{ds'_{ik}}{\pi}\frac{\rho_J(s'_{ik})G_J(s'_{ik})}
{s'_{ik}-s_{ik}},\eqno (25)$$

$$B_J(s_{ik})=\int\limits_{(m_i+m_k)^2}^{\infty}\hskip2mm
\frac{ds'_{ik}}{\pi}\frac{\rho_J(s'_{ik})G^2_J(s'_{ik})}
{s'_{ik}-s_{ik}},\eqno (26)$$

$$\rho_J (s_{ik})=\frac{(m_i+m_k)^2}{4\pi}
\left(\alpha_J\frac{s_{ik}}{(m_i+m_k)^2}
+\beta_J +\frac{\delta_J}{s_{ik}} \right)\times$$

$$\times\frac{\sqrt{(s_{ik}-(m_i+m_k)^2)(s_{ik}-(m_i-m_k)^2)}}
{s_{ik}}\, .\eqno (27)$$

Here $G_J$ is the diquark vertex function; $B_J(s_{ik})$, $\rho_J (s_{ik})$
are the Chew-Mandelstam function [24] and the phase space consequently.
$s_{ik}$ is the two-particle subenergy squared (i,k=1,2,3), $s$ is the
systems total energy squared.

For the state $J^p=\frac{3}{2} ^{-}$ of the (10,2) multiplet there are
three diquarks $J^p=1^+$, $1^+_S$, $1^-_S$. The coefficient of
Chew-Mandelstam function $\alpha_J$, $\beta_J$ and $\delta_J$ are given
in Table VII.

In the case in question the interacting quarks do not produce bound state,
then the integration in dispersion integrals is carried out from
$(m_i+m_k)^2$ to $\infty$.

All diagrams are classified over the last quark pair (Fig.1).

We use the diquark projectors. Then we consider the particle $\Sigma$
$\frac{3}{2} ^{-}$ of the $(10,2)$ multiplet again. This wave function
contain the contribution $u^1\downarrow u\uparrow s\uparrow$, which
include three diquarks: $u^1\downarrow u\uparrow$,\hskip 2mm
$u^1\downarrow s\uparrow$\hskip 2mm and \hskip 2mm $u\uparrow s\uparrow$.
The diquark projectors allow us to obtain the equations (28)-(30)
(with the definition $\rho_J(s_{ij})\equiv k_{ij}$).

$$k_{12}\left(\frac{a_1^0+a_0^1}{4}\left(u^1\downarrow u\uparrow s\uparrow
+u\uparrow u^1\downarrow s\uparrow\right)+\right.$$

$$\left. +\frac{a_1^0-a_0^1}{4}\left(
u^1\uparrow u\downarrow s\uparrow+u\downarrow u^1\uparrow s\uparrow
\right)\right)\hskip 2mm ,\eqno (28)$$

$$k_{13}\left(\frac{a_1^{0S}+a_0^{1S}+a_0^{0S}+a_1^{1S}}{8}
\left(u^1\downarrow u\uparrow s\uparrow
+s\uparrow u\uparrow u^1\downarrow\right)+\right.$$

$$\left. +\frac{a_1^{0S}-a_0^{1S}-a_0^{0S}+a_1^{1S}}{8}
\left(u^1\uparrow u\uparrow s\downarrow
+s\downarrow u\uparrow u^1\uparrow\right)+\right.$$

$$\left. +\frac{a_1^{0S}+a_0^{1S}-a_0^{0S}-a_1^{1S}}{8}
\left(s^1\downarrow u\uparrow u\uparrow
+u\uparrow u\uparrow s^1\downarrow\right)+\right.$$

$$\left. +\frac{a_1^{0S}-a_0^{1S}+a_0^{0S}-a_1^{1S}}{8}
\left(s^1\uparrow u\uparrow u\downarrow
+u\downarrow u\uparrow s^1\uparrow\right)
\right)\hskip 2mm ,\eqno (29)$$

$$k_{23}\left(\frac{a_1^{0S}}{2}\left(u^1\downarrow u\uparrow s\uparrow
+u^1\downarrow s\uparrow u\uparrow\right)
\right)\hskip 2mm .\eqno (30)$$

Then all members of wave function can be considered. And after the
groupping of these members we can obtain, for instance:

$$u^1\downarrow u\uparrow s\uparrow \left\{
k_{12}\frac{a_1^0+3a_0^1}{4}+k_{13}\frac{a_1^{0S}+3a_0^{1S}}{4}
+k_{23}\hskip 1mm a_1^{0S}\right\}\hskip 2mm .\eqno (31)$$

The left side of the diagram (Fig.2) corresponds to the
quark interactions. The rigth side of Fig.2
determines the zero approximation (first diagram) and the subsequent
pair interactions (second diagram). The contribution to
$u^1\downarrow u\uparrow s\uparrow$ is shown in the Fig.3.

If we group the same members, we obtain the system integral equations
for the $\Sigma$ state with the $J^p=\frac{3}{2} ^{-}$ of the $(10,2)$
multiplet:

$$\left\{
\begin{array}{l}
A_1^0(s,s_{12})=\lambda\, b_{1^+}(s_{12})L_{1^+}(s_{12})+
K_{1^+}(s_{12})\left[\frac{1}{4}A_1^{0S}(s,s_{13})+
\frac{3}{4}A_0^{1S}(s,s_{13})+\right.\\

\\
\hskip22mm \left.
+\frac{1}{4}A_1^{0S}(s,s_{23})+\frac{3}{4}A_0^{1S}(s,s_{23})
\right]\\

\\
A_1^{0S}(s,s_{13})=\lambda\, b_{1_S^+}(s_{13})L_{1_S^+}(s_{13})+
K_{1_S^+}(s_{13})\left[\frac{1}{2}A_1^0(s,s_{12})-
\frac{1}{4}A_1^{0S}(s,s_{12})+\right.\\

\\
\hskip22mm \left.
+\frac{3}{4}A_0^{1S}(s,s_{12})+\frac{1}{2}A_1^0(s,s_{23})-
\frac{1}{4}A_1^{0S}(s,s_{23})+\frac{3}{4}A_0^{1S}(s,s_{23})
\right] \\

\\
A_0^{1S}(s,s_{23})=\lambda\, b_{1_S^-}(s_{23})L_{1_S^-}(s_{23})+
K_{1_S^-}(s_{23})\left[\frac{1}{2}A_1^0(s,s_{12})+
\frac{1}{4}A_1^{0S}(s,s_{12})+\right.\\

\\
\hskip22mm \left.
+\frac{1}{4}A_0^{1S}(s,s_{12})+\frac{1}{2}A_1^0(s,s_{13})+
\frac{1}{4}A_1^{0S}(s,s_{13})+\frac{1}{4}A_0^{1S}(s,s_{13})
\right] \hskip2mm .\\
\end{array}
\right.\eqno (32)$$

Here function $L_J(s_{ik})$ has the form

$$L_J(s_{ik})=\frac{G_J(s_{ik})}{1-B_J(s_{ik})}.\eqno (33)$$

The integral operator $K_J (s_{ik})$ is:

$$K_J (s_{ik})=L_J(s_{ik})\, \int\limits_{(m_i+m_k)^2}^{\infty}\hskip2mm
\frac{ds'_{ik}}{\pi}\frac{\rho_J(s'_{ik})G_J(s'_{ik})}
{s'_{ik}-s_{ik}}\, \int\limits_{-1}^{1}\frac{dz}{2}\, .\eqno (34)$$

The function $b_J(s_{ik})$ is the truncate function of Chew-Mandelstam.
$z$ is the cosine of the angle between the relative momentum of particles
$i$ and $k$ in the intermediate state and the momentum of
particle $j$ in the final state, taken in the c.m. of the particles
$i$ and $k$. $\lambda$ is the current constant.

By analogy with the $\Sigma$ $\frac{3}{2} ^{-}$ $(10,2)$ state we obtain
the rescattering amplitudes of the three various quarks for all $P$-wave
states of the ${\bf (70,1^-)}$ multiplet which satisfy the system of
integral equations (Appendix B).

\vskip 5mm
{\bf 4. The reduced equations of ${\bf (70,1^-)}$ multiplet.}
\vskip 5mm

Let us extract two-particle singularities in $A_J(s,s_{ik})$:

$$A_J(s,s_{ik})=\frac{\alpha_J(s,s_{ik})b_J(s_{ik})G_J(s_{ik})}
{1-B_J(s_{ik})},\eqno (35)$$

\noindent
$\alpha_J(s,s_{ik})$ is the reduced amplitude. Accordingly to this,
all integral equations can be rewritten using the reduced
amplitudes. For instance, one consider the first equation of system
for the $\Sigma$ $J^p=\frac{3}{2}^-$ of the $(10,2)$ multiplet:

$$\alpha_1^0 (s,s_{12})=\lambda+\frac{1}{b_{1^+}(s_{12})}
\, \int\limits_{(m_1+m_2)^2}^{\Lambda_{1^+}(1,2)}\hskip2mm
\frac{ds'_{12}}{\pi}\,\frac{\rho_{1^+}(s'_{12})G_{1^+}(s'_{12})}
{s'_{12}-s_{12}}\times$$

$$\times\int\limits_{-1}^{1}\frac{dz}{2}\,
\left(
\frac{G_{1_S^+}(s'_{13})b_{1_S^+}(s'_{13})}{1-B_{1_S^+}(s'_{13})}
\,\frac{1}{2}\,\alpha_1^{0S}(s,s'_{13})+
\frac{G_{1_S^-}(s'_{13})b_{1_S^-}(s'_{13})}{1-B_{1_S^-}(s'_{13})}
\,\frac{3}{2}\,\alpha_0^{1S}(s,s'_{13})
\right).\eqno (36)$$

The connection between $s'_{12}$ and $s'_{13}$ is [25]:

$$s'_{13}=m_1^2+m_3^2-\frac{\left(s'_{12}+m_3^2-s\right)
\left(s'_{12}+m_1^2-m_2^2\right)}{2s'_{12}}\pm$$

$$\pm\frac{z}{2s'_{12}}\times\sqrt{\left(s'_{12}-(m_1+m_2)^2\right)
\left(s'_{12}-(m_1-m_2)^2\right)}\times$$

$$\times\sqrt{\left(s'_{12}-(\sqrt{s}+m_3)^2\right)
\left(s'_{12}-(\sqrt{s}-m_3)^2\right)}\, .\eqno (37)$$

The formula for $s'_{23}$ is similar to (37) with $z$ replaced by $-z$.
Thus $A_1^{0S}(s,s'_{13})+A_1^{0S}(s,s'_{23})$ must be replaced by
$2A_1^{0S}(s,s'_{13})$. $\Lambda_J(i,k)$ is the cutoff at the large
value of $s_{ik}$, which determines the contribution from small distances.

The construction of the approximate solution of the (36) is based on
the extraction of the leading singularities which are close to the
region $s_{ik}=(m_i+m_k)^2$ [25]. Amplitudes with different number of
rescattering have the following structure of singularities. The main
singularities in $s_{ik}$ are from pair rescattering of the particles
$i$ and $k$. First of all there are threshold square root singularities.
Pole singularities are also possible which correspond to the bound
states. The diagrams in Fig.2 apart from two-particle singularities
have their own specific triangle singularities. Such classification
allows us to search the approximate solution of (36) by taking into
account some definite number of leading singularities and neglecting all
the weaker ones.

We consider the approximation, which corresponds to the single interaction
of all three particles (two-particle and triangle singularities) and
neglecting all the weaker ones.

The functions $\alpha_J(s,s_{ik})$ are the smooth functions of $s_{ik}$
as compared with the singular part of the amplitude, hence it can be
expanded in a series in the singulary point and only the first term of
this series should be employed further. As $s_0$ it is convenient to
take the middle point of physical region of Dalitz-plot in which $z=0$.
In this case we get from (37)
$s_{ik}=s_0=\frac{s+m_1^2+m_2^2+m_3^2}{m_{12}^2+m_{13}^2+m_{23}^2}$,
where $m_{ik}=\frac{m_i+m_k}{2}$. We define the functions
$\alpha_J(s,s_{ik})$ and $b_J(s_{ik})$ at the point $s_0$. Such a choice
of point $s_0$ allows us to replace integral equations (36) by the
algebraic equations for the state $\Sigma$ with $J^p=\frac{3}{2}^-$
of the $(10,2)$ multiplet:

$$\left\{
\begin{array}{l}
\alpha_1^0(s,s_0)=\lambda+\frac{1}{2}\hskip2mm\alpha_1^{0S}(s,s_0)
\hskip2mm I_{1^+ 1^+_S}(s,s_0)
\hskip2mm\frac{b_{1^+_s}(s_0)}{b_{1^+}(s_0)}\hskip40mm 1^+\\
\hskip25mm
+\frac{3}{2}\hskip2mm\alpha_0^{1S}(s,s_0)\hskip2mm I_{1^+ 1^-_S}(s,s_0)
\hskip2mm\frac{b_{1^-_s}(s_0)}{b_{1^+}(s_0)}\\
\alpha_1^{0S}(s,s_0)=\lambda
+\alpha_1^0(s,s_0)\hskip2mm I_{1^+_S 1^+}(s,s_0)
\hskip2mm\frac{b_{1^+}(s_0)}{b_{1^+_s}(s_0)}\hskip45mm 1^+_S\\
\hskip25mm
-\frac{1}{2}\hskip2mm\alpha_1^{0S}(s,s_0)
\hskip2mm I_{1^+_S 1^+_S}(s,s_0)
+\frac{3}{2}\hskip2mm\alpha_0^{1S}(s,s_0)\hskip2mm I_{1^+_S 1^-_S}(s,s_0)
\hskip2mm\frac{b_{1^-_s}(s_0)}{b_{1^+_s}(s_0)}\\
\alpha_0^{1S}(s,s_0)=\lambda+\alpha_1^0(s,s_0)\hskip2mm I_{1^-_S 1^+}(s,s_0)
\hskip2mm\frac{b_{1^+}(s_0)}{b_{1^-_s}(s_0)}\hskip45mm 1^-_S\\
\hskip25mm
+\frac{1}{2}\hskip2mm\alpha_1^{0S}(s,s_0)
\hskip2mm I_{1^-_S 1^+_S}(s,s_0)
\hskip2mm\frac{b_{1^+_s}(s_0)}{b_{1^-_s}(s_0)}
+\frac{1}{2}\hskip2mm\alpha_0^{1S}(s,s_0)\hskip2mm I_{1^-_S 1^-_S}(s,s_0)
\hskip2mm .\\
\end{array} \right.\eqno (38)$$

Here the reduced amplitudes for the diquarks $1^+$, $1^+_s$, $1^-_s$
are given.

The function $I_{J_1 J_2}(s,s_0)$ takes into account singularity which
corresponds to the simultaneous vanishing of all propagators in the
triangle diagrams.

$$I_{J_1 J_2}(s,s_0)=\int\limits_{(m_i+m_k)^2}^{\Lambda_{J_1}}\hskip2mm
\frac{ds'_{ik}}{\pi}\frac{\rho_{J_1}(s'_{ik})G^2_{J_1}(s'_{ik})}
{s'_{ik}-s_{ik}}\, \int\limits_{-1}^{1}\frac{dz}{2}\,
\frac{1}{1-B_{J_2}(s_{ij})}\eqno (39)$$

The $G_J(s_{ik})$ functions have the smooth dependence from energy
$s_{ik}$ [23, 26] therefore we suggest them as constants.
The parameters of model: $\lambda_J$ cut off parameter, $g_J$ vertex
constants are chosen dimensionless.

$$g_J=\frac{m_i+m_k}{2\pi}G_J , \,\,\, \lambda_J=\frac{4\Lambda_J}
{(m_i+m_k)^2} .\eqno (40)$$

Here $m_i$ and $m_k$ are quark masses in the intermediate state of the quark
loop. Dimensionless parameters $g_J$ and $\lambda_J$ are supposed to be
the constants independent of the quark interaction type. We calculate
the system equations and can determine the mass values of the
$\Sigma$ $J^p=\frac{3}{2}^-$ $(10,2)$. We calculate a pole in $s$ which
corresponds to the bound state of the three quarks.

By analogy with $\Sigma$-hyperon we obtain the system equations for the
reduced amplitudes for all particles $\bf{(70,1^-)}$ multiplets.

\vskip 5mm
{\bf 5. Calculation results.}
\vskip 5mm

The quark masses ($m_u=m_d=m$ and $m_s$) are not fixed. In order to fix
$m$ and $m_s$, in any way we assume $m=\frac{1}{3}m_{\Delta}(1.232)$ and
$m=\frac{1}{3}m_{\Omega}(1.672)$ i.e. the quark masses are $m=0.410\, GeV$
and $m_s=0.557\, GeV$.

The $S$-wave baryon mass spectra are obtained in good agreement with the
experimental data. When we research the excited states the
confinement potential can not be neglected. In our case the confinement
potential is imitated by the simple increasing of constituent quark
masses [27]. The shift of quark mass (parameter $\Delta=160\, MeV$)
effectively takes into account the changing of the confinement potential.
We have shown that inclusion of only gluon exchange does not lead
to the appearance of bound states corresponding to the baryons in the
$P$-wave. The use of mass shift $\Delta$ is possible to obtain the
mass spectra of $P$-wave baryons. The similar result for the $P$-wave
mesons was obtained [27].

In the case considered the same parameters $\Delta$ for the
$u, d, s$-quarks are choosen. Then the quark masses $m_u=m_d=0.570\, GeV$
and $m_s=0.717\, GeV$ are given.

In our model the four parameters are used: gluon coupling constant
$g_+$ and $g_-$ for various parity, cutoff energy parameters
$\lambda$, $\lambda _s$ for the nonstrange and strange diquarks.

The parameters $g_+ =0.69$, $g_- =0.30$, $\lambda=14.5$, $\lambda _s=11.2$
have been determined by the baryon masses:
$M_{N \frac{1}{2}^- (8,4)}=1.650\, GeV$,\,
$M_{N \frac{1}{2}^- (8,2)}=1.535\, GeV$,\,
$M_{N \frac{3}{2}^- (8,2)}=1.520\, GeV$, and
$M_{\Lambda \frac{3}{2}^- (1,2)}=1.520\, GeV$.
In the table I-VI we present the masses of the nonstrange and strange
resonances belonging to the ${\bf (70,1^-)}$ multiplet obtained using
the fit of the experimental values [28].

The ${\bf (70,1^-)}$ multiplet include $210$ particles, only $30$ baryons
have different masses. We have predicted $15$ masses of baryons.

In the framework of the proposed approximate method of solving the
relativistic three-particle problem, we have obtained a satisfactory
spectrum of $P$-wave baryons. The important problem is the mixing
of $P$-wave baryons and the five quark system (cryptoexotic baryons [29]
or hybrid baryons [30]).

\vskip 5mm
{\bf 6. Conclusion.}
\vskip 5mm

Many potential model calculations, which retain the one-gluon exchange
picture of the quark-quark interactions have gone beyond Isgur and Karl
model (see the reviews [31, 32]) for the spectrum and wave functions of
baryons in attempt to correct the flaws in the nonrelativistic model [33].
The relativized quark model applied to meson spectroscopy by Capstick and
Isgur is similarly considered [33].

In the papers [19, 20] the relativistic generalization of Faddeev equations
in the framework of dispersion relation are constructed. We calculated
the $S$-wave baryon masses using the method based on the extraction of
leading singularities of the amplitude. The behavior of electromagnetic
form factor of the nucleon and hyperon in the region of low and
intermediate momentum transfers is determined by [19, 34].

In the framework of the dispersion relation approach the charge radii
of $S$-wave baryon multiplets with $J^p=\frac{1}{2}^+$ are calculated.

In the present paper the relativistic description of three particles
amplitudes of $P$-wave baryons are considered. We take into account
the $u, d, s$-quarks. The mass spectrum of nonstrange and strange states
of multiplet ${\bf (70,1^-)}$ are calculated. We use only four parameters
for the calculation of $30$ baryon masses. We take into account the mass
shift of $u, d, s$ quarks which allows us to obtain the $P$-wave baryon
bound states.

Recently, the mass spectrum nonstrange baryons of ${\bf (70,1^-)}$
using $1/N_c$ expansion are calculated [13].

In this paper the orbital-spin-flavor wave function is constructed.
The authors solve the problem of difference between the ground state
and the excited states for the ${\bf (70,1^-)}$ multiplet.

We also use the orbital-spin-flavor wave function for the construction
of integral equations. It allows to calculate the mass spectra for
all baryons ${\bf (70,1^-)}$ multiplet.

We can see that the masses of $P$-wave baryons with $J^p=\frac{1}{2}^-$
are heavier than the masses of states with $J^p=\frac{3}{2}^-$ and
$J^p=\frac{5}{2}^-$. This conclusion contradicts to the result of
nonrelativistic quark models [31]. The exceptions are the
masses of lowest $\Lambda$-baryons with $J^p=\frac{1}{2}^-$ and
$J^p=\frac{3}{2}^-$.

This model shows some improvements relative to the nonrelativistic model,
largely because it does not contain the possibility to separately fit the
negative-parity and positive-parity states present in the nonrelativistic
model.

We considered the relativistic three-quark equations and calculated the
mass spectrum of ${\bf (70,1^-)}$ baryon multiplet.

The following problem of excited baryon description is the mass spectrum
of ${\bf (70,L^+)}$, $L^+=0, 2$ baryons.

The authors would like to thank T. Barnes, N.N. Nikolaev, Fl. Stancu
for useful discussions. The work was carried with the support of the
Russion Ministry of Education (grant 2.1.1.68.26).

\newpage
{\bf Appendix A. The ${\bf P}$-wave baryon wave functions.}
\vskip 5mm
{\bf A1. The wave functions of ${\bf (10,2)}$ decuplet.}
\vskip 5mm

We considered this decuplet in the Section 2. The totally symmetric
$SU(6)\times O(3)$ wave function for each decuplet particle has the
following form:

$$\varphi=\frac{1}{\sqrt{2}}\left(
\varphi_{MA}^{SU(6)}\varphi_{MA}^{O(3)}+
\varphi_{MS}^{SU(6)}\varphi_{MS}^{O(3)}
\right)=\frac{1}{\sqrt{2}}\varphi_{S}^{SU(3)}\left(
\varphi_{MA}^{SU(2)}\varphi_{MA}^{O(3)}+
\varphi_{MS}^{SU(2)}\varphi_{MS}^{O(3)}
\right).\eqno (41)$$

The functions $\varphi_{MA}^{SU(2)}$, $\varphi_{MS}^{SU(2)}$,
$\varphi_{MA}^{O(3)}$, $\varphi_{MS}^{O(3)}$ are:

$$\varphi_{MA}^{SU(2)}=\frac{1}{\sqrt{2}}\left(
\uparrow \downarrow \uparrow-\downarrow \uparrow \uparrow
\right),\quad
\varphi_{MS}^{SU(2)}=\frac{1}{\sqrt{6}}\left(
\uparrow \downarrow \uparrow+\downarrow \uparrow \uparrow-
2\uparrow \uparrow \downarrow
\right),\eqno (42)$$

$$\varphi_{MA}^{O(3)}=\frac{1}{\sqrt{2}}\left(010-100\right),\quad
\varphi_{MS}^{O(3)}=\frac{1}{\sqrt{6}}\left(
010+100-2\cdot 001\right).\eqno (43)$$

For the $\Sigma^{+}$-hyperon $SU(3)$-function is:

$$\varphi_{S}^{SU(3)}=\frac{1}{\sqrt{3}}\left(
usu+suu+uus
\right).\eqno (44)$$

Then one obtain the $SU(6)\times O(3)$-function of $\Sigma$ the
$(10,2)$ multiplet:

$$\varphi_{\Sigma^{+}(10,2)}=\frac{\sqrt{6}}{18}\left(
2\{u^1\downarrow u\uparrow s\uparrow\}+
\{s^1\downarrow u\uparrow u\uparrow\}-\right.$$
$$\left.-\{u^1\uparrow u\downarrow s\uparrow\}-
\{u^1\uparrow u\uparrow s\downarrow \}-
\{s^1\uparrow u\uparrow u\downarrow \}\right).\eqno (45)$$

In the case of $\Delta$ the $SU(6)\times O(3)$ wave functions are given:

$$\varphi_{\Delta^{++}(10,2)}=\frac{\sqrt{2}}{6}\left(
\{u^1\downarrow u\uparrow u\uparrow\}
-\{u^1\uparrow u\uparrow u\downarrow\}
\right).\eqno (46)$$

The replacement by $u\leftrightarrow s$ or $d\leftrightarrow s$ allows
us to obtain the $\Xi$ wave function using the $\Sigma$ wave function.
The $\Omega^{-}$ function is similar to the $\Delta$ wave function.

\vskip 5mm
{\bf A2. The wave functions of ${\bf (8,2)}$ octet.}
\vskip 5mm
The wave functions of octet $\frac{3}{2} ^{-} ,\frac{1}{2} ^{-}$
$(8,2)$ multiplet are constructed as:

$$\varphi=\frac{1}{\sqrt{2}}\left(
\varphi_{MA}^{SU(6)}\varphi_{MA}^{O(3)}+
\varphi_{MS}^{SU(6)}\varphi_{MS}^{O(3)}
\right),\eqno (47)$$

\noindent
here

$$\varphi_{MA}^{SU(6)}=\frac{1}{\sqrt{2}}\left(
\varphi_{MS}^{SU(3)}\varphi_{MA}^{SU(2)}+
\varphi_{MA}^{SU(3)}\varphi_{MS}^{SU(2)}
\right),\eqno (48)$$

$$\varphi_{MS}^{SU(6)}=\frac{1}{\sqrt{2}}\left(
-\varphi_{MS}^{SU(3)}\varphi_{MS}^{SU(2)}+
\varphi_{MA}^{SU(3)}\varphi_{MA}^{SU(2)}
\right).\eqno (49)$$

In the case of $\Sigma^{+}$ the $SU(3)$ wave functions
$\varphi_{MS}^{SU(3)}$ and $\varphi_{MA}^{SU(3)}$ have the following form:

$$\varphi_{MS}^{SU(3)}=\frac{1}{\sqrt{6}}\left(
usu+suu-2uus\right),\quad
\varphi_{MA}^{SU(3)}=\frac{1}{\sqrt{2}}\left(
usu-suu\right).\eqno (50)$$

Then we can obtain the symmetric wave function for $\Sigma^{+}$:

$$\varphi_{\Sigma^{+} (8,2)}=\frac{\sqrt{6}}{18}\left(
2\{u^1\uparrow u\downarrow s\uparrow\}+
\{s^1\downarrow u\uparrow u\uparrow\}-
\{u^1\uparrow u\uparrow s\downarrow\}-\right.$$

$$\left.-\{u^1\downarrow u\uparrow s\uparrow\}-
\{s^1\uparrow u\uparrow u\downarrow\}
\right).\eqno (51)$$

By analogy with the $\Sigma^{+}$ we calculate the $\Sigma^{0}$- and
$\Sigma^{-}$ wave functions.

The nucleon $\frac{3}{2} ^{-} ,\frac{1}{2} ^{-}$ $(8,2)$ wave functions
are obtained with the replacement by $s\leftrightarrow d$ in $\Sigma^{+}$,
and $\Xi^{0}$ with replacement by $u\leftrightarrow s$ in $\Sigma^{+}$.

The $\Lambda^{0}$ $SU(3)$ wave functions $\varphi_{MS}^{SU(3)}$ and
$\varphi_{MA}^{SU(3)}$ are given:

$$\varphi_{MS}^{SU(3)}=\frac{1}{2}\left(
dsu-usd+sdu-sud
\right),\eqno (52)$$

$$\varphi_{MA}^{SU(3)}=\frac{\sqrt{3}}{6}\left(
sdu-sud+usd-dsu-2dus+2uds
\right).\eqno (53)$$

Then the symmetric $SU(6)\times O(3)$ wave function for $\Lambda^{0}$
$\frac{3}{2} ^{-} ,\frac{1}{2} ^{-}$ can be considered as:

$$\varphi_{\Lambda^{0} (8,2)}=\frac{1}{6}\left(
\{u^1\uparrow d\uparrow s\downarrow\}-
\{u^1\downarrow d\uparrow s\uparrow\}-
\{d^1\uparrow u\uparrow s\downarrow\}+
\right.$$
$$\left.
+\{d^1\downarrow u\uparrow s\uparrow\}-
\{s^1\uparrow u\uparrow d\downarrow\}+
\{s^1\uparrow u\downarrow d\uparrow\}
\right).\eqno (54)$$

Here we can see that the contribution $u^1\uparrow s\uparrow d\downarrow$
is absent in the wave function of $\Lambda^{0}$
$\frac{3}{2} ^{-} ,\frac{1}{2} ^{-}$ $(8,2)$. One can conclude that the
diquark $1^+$ do not include into the corresponding amplitude.
\vskip 5mm
{\bf A3. The wave function of ${\bf (8,4)}$ octet.}
\vskip 5mm
By analogy with the cases $(10,2)$ and $(8,2)$ we can calculate the
$(8,4)$ octet wave functions:

$$\varphi=\frac{1}{\sqrt{2}}\left(
\varphi_{MA}^{SU(6)}\varphi_{MA}^{O(3)}+
\varphi_{MS}^{SU(6)}\varphi_{MS}^{O(3)}
\right),\eqno (55)$$

\noindent
here

$$\varphi_{MA}^{SU(6)}=
\varphi_{MA}^{SU(3)}\varphi_{S}^{SU(2)},\quad
\varphi_{MS}^{SU(6)}=
\varphi_{MS}^{SU(3)}\varphi_{S}^{SU(2)}.\eqno (56)$$

$SU(2)$ wave function is totally symmetric:

$$\varphi_{S}^{SU(2)}=\uparrow\uparrow\uparrow ,\eqno (57)$$

\noindent
$\varphi_{MS}^{SU(3)}$ and $\varphi_{MA}^{SU(3)}$ similar to one of the
$(8,2)$ multiplet.

For the $\Sigma^{+}$ $\frac{3}{2} ^{-} ,\frac{1}{2} ^{-}$ of $(8,4)$
multiplet one have:

$$\varphi_{\Sigma^{+} (8,4)}=\frac{\sqrt{2}}{6}\left(
\{s^1\uparrow u\uparrow u\uparrow\}-
\{u^1\uparrow u\uparrow s\uparrow\}
\right).\eqno (58)$$

For the nucleon $N$ we can replace by $s\rightarrow d$ in $\Sigma^{+}$,
and $u\leftrightarrow s$ for the $\Xi^{0}$.

We obtain the wave function of the $\Lambda^{0}$ $(8,4)$:

$$\varphi_{\Lambda^{0} (8,4)}=\frac{\sqrt{3}}{6}\left(
-\{u^1\uparrow d\uparrow s\uparrow\}+
\{d^1\uparrow u\uparrow s\uparrow\}
\right).\eqno (59)$$

\vskip 5mm
{\bf A4. The ${\bf (1,2)}$ singlet wave function.}
\vskip 5mm
We must use the other method if we consider the
$\frac{3}{2} ^{-} ,\frac{1}{2} ^{-}$ $(1,2)$ singlet $\Lambda^{0}_{1}$.
We can use the totally symmetric $SU(6)\times O(3)$ wave function in the
form:

$$\varphi=\varphi_{A}^{SU(3)}\varphi_{A}^{SU(2)\times O(3)},\eqno (60)$$

\noindent
here

$$\varphi_{A}^{SU(3)}=\frac{1}{\sqrt{6}}\left(
sdu-sud+usd-dsu+dus-uds
\right),\eqno (61)$$

$$\varphi_{A}^{SU(2)\times O(3)}=\frac{1}{\sqrt{2}}\left(
\varphi_{MS}^{SU(2)}\varphi_{MA}^{O(3)}-
\varphi_{MA}^{SU(2)}\varphi_{MS}^{O(3)}
\right).\eqno (62)$$

As result we obtain:

$$\varphi_{\Lambda^{0}_{1} (1,2)}=\frac{\sqrt{3}}{6}\left(
-\{u^1\uparrow d\uparrow s\downarrow\}+
\{u^1\uparrow d\downarrow s\uparrow\}+
\{d^1\uparrow u\uparrow s\downarrow\}-\right.$$
$$\left.
-\{d^1\uparrow u\downarrow s\uparrow\}-
\{s^1\uparrow u\uparrow d\downarrow\}+
\{s^1\uparrow u\downarrow d\uparrow\}
\right).\eqno (63)$$

\vskip 5mm
{\bf Appendix B. The integral equations for the ${\bf (70,1^-)}$ multiplet.}
\vskip 5mm
{\bf B1. ${\bf (10,2)}$ multiplet.}
\vskip 5mm

We can represent the equations for the $\frac{3}{2}^-$ $(10,2)$ multiplet,
which is determined by the projection of orbital moment $l_z=+1$.
We consider the following states:

$\Delta$ $\frac{3}{2} ^{-}$:

$$\left\{
\begin{array}{l}
A_1^0(s,s_{12})=\lambda\, b_{1^+}(s_{12})L_{1^+}(s_{12})+
K_{1^+}(s_{12})\left[\frac{1}{4}A_1^0(s,s_{13})+\frac{3}{4}A_0^1(s,s_{13})+
\right.\\

\\
\hskip22mm \left.
+\frac{1}{4}A_1^0(s,s_{23})+\frac{3}{4}A_0^1(s,s_{23})
\right]\\

\\
A_0^1(s,s_{13})=\lambda\, b_{1^-}(s_{13})L_{1^-}(s_{13})+
K_{1^-}(s_{13})\left[\frac{3}{4}A_1^0(s,s_{12})+\frac{1}{4}A_0^1(s,s_{12})+
\right.\\

\\
\hskip22mm \left.
+\frac{3}{4}A_1^0(s,s_{23})+\frac{1}{4}A_0^1(s,s_{23})
\right] \hskip2mm .\\
\end{array}
\right.\eqno (64)$$

\newpage
$\Sigma$ $\frac{3}{2} ^{-}$:

$$\left\{
\begin{array}{l}
A_1^0(s,s_{12})=\lambda\, b_{1^+}(s_{12})L_{1^+}(s_{12})+
K_{1^+}(s_{12})\left[\frac{1}{4}A_1^{0S}(s,s_{13})+
\frac{3}{4}A_1^{0S}(s,s_{13})+\right.\\

\\
\hskip22mm \left.
+\frac{1}{4}A_1^{0S}(s,s_{23})+\frac{3}{4}A_0^{1S}(s,s_{23})
\right]\\

\\
A_1^{0S}(s,s_{13})=\lambda\, b_{1_S^+}(s_{13})L_{1_S^+}(s_{13})+
K_{1_S^+}(s_{13})\left[\frac{1}{2}A_1^0(s,s_{12})-
\frac{1}{4}A_1^{0S}(s,s_{12})+\right.\\

\\
\hskip22mm \left.
+\frac{3}{4}A_0^{1S}(s,s_{12})+\frac{1}{2}A_1^0(s,s_{23})-
\frac{1}{4}A_1^{0S}(s,s_{23})+\frac{3}{4}A_0^{1S}(s,s_{23})
\right] \\

\\
A_0^{1S}(s,s_{23})=\lambda\, b_{1_S^-}(s_{23})L_{1_S^-}(s_{23})+
K_{1_S^-}(s_{23})\left[\frac{1}{2}A_1^0(s,s_{12})+
\frac{1}{4}A_1^{0S}(s,s_{12})+\right.\\

\\
\hskip22mm \left.
+\frac{1}{4}A_0^{1S}(s,s_{12})+\frac{1}{2}A_1^0(s,s_{13})+
\frac{1}{4}A_1^{0S}(s,s_{13})+\frac{1}{4}A_0^{1S}(s,s_{13})
\right] \hskip2mm .\\
\end{array}
\right.\eqno (65)$$

The system integral equations for the $\Xi$ $\frac{3}{2} ^{-}$ are similar
to $\Sigma$ $\frac{3}{2} ^{-}$ with the replacement by $u\leftrightarrow s$:

$$\left\{
\begin{array}{l}
A_1^{0SS}(s,s_{12})=\lambda\, b_{1^+_{SS}}(s_{12})L_{1^+_{SS}}(s_{12})+
K_{1^+_{SS}}(s_{12})\left[\frac{1}{4}A_1^{0S}(s,s_{13})+
\frac{3}{4}A_1^{0S}(s,s_{13})+\right.\\

\\
\hskip22mm \left.
+\frac{1}{4}A_1^{0S}(s,s_{23})+\frac{3}{4}A_0^{1S}(s,s_{23})
\right]\\

\\
A_1^{0S}(s,s_{13})=\lambda\, b_{1_S^+}(s_{13})L_{1_S^+}(s_{13})+
K_{1_S^+}(s_{13})\left[\frac{1}{2}A_1^{0SS}(s,s_{12})-
\frac{1}{4}A_1^{0S}(s,s_{12})+\right.\\

\\
\hskip22mm \left.
+\frac{3}{4}A_0^{1S}(s,s_{12})+\frac{1}{2}A_1^{0SS}(s,s_{23})-
\frac{1}{4}A_1^{0S}(s,s_{23})+\frac{3}{4}A_0^{1S}(s,s_{23})
\right] \\

\\
A_0^{1S}(s,s_{23})=\lambda\, b_{1_S^-}(s_{23})L_{1_S^-}(s_{23})+
K_{1_S^-}(s_{23})\left[\frac{1}{2}A_1^{0SS}(s,s_{12})+
\frac{1}{4}A_1^{0S}(s,s_{12})+\right.\\

\\
\hskip22mm \left.
+\frac{1}{4}A_0^{1S}(s,s_{12})+\frac{1}{2}A_1^{0SS}(s,s_{13})+
\frac{1}{4}A_1^{0S}(s,s_{13})+\frac{1}{4}A_0^{1S}(s,s_{13})
\right] \hskip2mm .\\
\end{array}
\right.\eqno (66)$$

Then we determine the system integral equations for
$\Omega$ $\frac{3}{2} ^{-}$ by analogy with $\Delta$ (the replacement
$u \rightarrow s$).

The multiplet $\frac{1}{2}^-$ for $(10,2)$ can be obtain if we use the
projection of orbital moment $l_z=0$. The equations of this multiplet
are obtained by the equations, which correspond to the particles
of multiplet $\frac{3}{2}^-$ $(10,2)$ (we replace diquark amplitudes):
$1^- \to 0^-$ \, $(A_0^1 \to A_0^{-0})$, \, \, $1^-_S \to 0^-_S$ \,
$(A_0^{1S} \to A_0^{-0S})$.

\newpage
$\Delta$ $\frac{1}{2} ^{-}$ $(10,2)$:

$$\left\{
\begin{array}{l}
A_1^0(s,s_{12})=\lambda\, b_{1^+}(s_{12})L_{1^+}(s_{12})+
K_{1^+}(s_{12})\left[\frac{1}{4}A_1^0(s,s_{13})+
\frac{3}{4}A_0^{-0}(s,s_{13})+\right.\\

\\
\hskip22mm \left.
+\frac{1}{4}A_1^0(s,s_{23})+\frac{3}{4}A_0^{-0}(s,s_{23})
\right]\\

\\
A_0^{-0}(s,s_{13})=\lambda\, b_{0^-}(s_{13})L_{0^-}(s_{13})+
K_{0^-}(s_{13})\left[\frac{3}{4}A_1^0(s,s_{12})+
\frac{1}{4}A_0^{-0}(s,s_{12})+\right.\\

\\
\hskip22mm \left.
+\frac{3}{4}A_1^0(s,s_{23})+\frac{1}{4}A_0^{-0}(s,s_{23})
\right] \hskip2mm .\\
\end{array}
\right.\eqno (67)$$

$\Sigma$ $\frac{1}{2} ^{-}$ $(10,2)$:

$$\left\{
\begin{array}{l}
A_1^0(s,s_{12})=\lambda\, b_{1^+}(s_{12})L_{1^+}(s_{12})+
K_{1^+}(s_{12})\left[\frac{1}{4}A_1^{0S}(s,s_{13})+
\frac{3}{4}A_1^{-0S}(s,s_{13})+\right.\\

\\
\hskip22mm \left.
+\frac{1}{4}A_1^{0S}(s,s_{23})+\frac{3}{4}A_0^{-0S}(s,s_{23})
\right]\\

\\
A_1^{0S}(s,s_{13})=\lambda\, b_{1_S^+}(s_{13})L_{1_S^+}(s_{13})+
K_{1_S^+}(s_{13})\left[\frac{1}{2}A_1^0(s,s_{12})-
\frac{1}{4}A_1^{0S}(s,s_{12})+\right.\\

\\
\hskip22mm \left.
+\frac{3}{4}A_0^{-0S}(s,s_{12})+\frac{1}{2}A_1^0(s,s_{23})-
\frac{1}{4}A_1^{0S}(s,s_{23})+\frac{3}{4}A_0^{-0S}(s,s_{23})
\right] \\

\\
A_0^{-0S}(s,s_{23})=\lambda\, b_{0_S^-}(s_{23})L_{0_S^-}(s_{23})+
K_{0_S^-}(s_{23})\left[\frac{1}{2}A_1^0(s,s_{12})+
\frac{1}{4}A_1^{0S}(s,s_{12})+\right.\\

\\
\hskip22mm \left.
+\frac{1}{4}A_0^{-0S}(s,s_{12})+\frac{1}{2}A_1^0(s,s_{13})+
\frac{1}{4}A_1^{0S}(s,s_{13})+\frac{1}{4}A_0^{-0S}(s,s_{13})
\right] \hskip2mm .\\
\end{array}
\right.\eqno (68)$$

By analogy with the case $\frac{3}{2} ^{-}$ $(10,2)$, the system integral
equations for $\Xi$ $\frac{1}{2} ^{-}$ $(10,2)$ is similar to the case
$\Sigma$ $\frac{1}{2} ^{-}$ $(10,2)$ with the replacement by
$u\leftrightarrow s$ and the system equations for $\Omega$
$\frac{1}{2} ^{-}$ $(10,2)$ is similar to the system for $\Delta$
$\frac{1}{2} ^{-}$ $(10,2)$ with the replacement by $u \rightarrow s$.

\newpage
{\bf B2. ${\bf (8,2)}$ multiplet.}
\vskip 5mm

We determine the equations of multiplet $\frac{3}{2}^-$ $(8,2)$ with
$l_z=+1$.

$N$ $\frac{3}{2} ^{-}$ $(8,2)$:

$$\left\{
\begin{array}{l}
A_1^0(s,s_{12})=\lambda\, b_{1^+}(s_{12})L_{1^+}(s_{12})+
K_{1^+}(s_{12})\left[-\frac{1}{8}A_1^0(s,s_{13})+\frac{3}{8}A_0^1(s,s_{13})+
\right.\\

\\
\hskip22mm
+\frac{3}{8}A_0^0(s,s_{13})+\frac{3}{8}A_1^1(s,s_{13})
-\frac{1}{8}A_1^0(s,s_{23})+\frac{3}{8}A_0^1(s,s_{23})+\\

\\
\hskip22mm \left.
+\frac{3}{8}A_0^0(s,s_{23})+\frac{3}{8}A_1^1(s,s_{23})
\right]\\

\\
A_0^1(s,s_{13})=\lambda\, b_{1^-}(s_{13})L_{1^-}(s_{13})+
K_{1^-}(s_{13})\left[\frac{3}{8}A_1^0(s,s_{12})-\frac{1}{8}A_0^1(s,s_{12})+
\right.\\

\\
\hskip22mm
+\frac{3}{8}A_0^0(s,s_{12})+\frac{3}{8}A_1^1(s,s_{12})
+\frac{3}{8}A_1^0(s,s_{23})-\frac{1}{8}A_0^1(s,s_{23})+\\

\\
\hskip22mm \left.
+\frac{3}{8}A_0^0(s,s_{23})+\frac{3}{8}A_1^1(s,s_{23})
\right]\\

\\
A_0^0(s,s_{23})=\lambda\, b_{0^+}(s_{23})L_{0^+}(s_{23})+
K_{0^+}(s_{23})\left[\frac{3}{8}A_1^0(s,s_{12})+\frac{3}{8}A_0^1(s,s_{12})-
\right.\\

\\
\hskip22mm
-\frac{1}{8}A_0^0(s,s_{12})+\frac{3}{8}A_1^1(s,s_{12})
+\frac{3}{8}A_1^0(s,s_{13})+\frac{3}{8}A_0^1(s,s_{13})-\\

\\
\hskip22mm \left.
-\frac{1}{8}A_0^0(s,s_{13})+\frac{3}{8}A_1^1(s,s_{13})
\right]\\

\\
A_1^1(s,s_{13})=\lambda\, b_{2^-}(s_{13})L_{2^-}(s_{13})+
K_{2^-}(s_{13})\left[\frac{3}{8}A_1^0(s,s_{12})+\frac{3}{8}A_0^1(s,s_{12})+
\right.\\

\\
\hskip22mm
+\frac{3}{8}A_0^0(s,s_{12})-\frac{1}{8}A_1^1(s,s_{12})
+\frac{3}{8}A_1^0(s,s_{23})+\frac{3}{8}A_0^1(s,s_{23})+\\

\\
\hskip22mm \left.
+\frac{3}{8}A_0^0(s,s_{23})-\frac{1}{8}A_1^1(s,s_{23})
\right]\hskip2mm .\\
\end{array}
\right.\eqno (69)$$

\newpage
$\Sigma$ $\frac{3}{2} ^{-}$ $(8,2)$:

$$\left\{
\begin{array}{l}
A_1^0(s,s_{12})=\lambda\, b_{1^+}(s_{12})L_{1^+}(s_{12})+
K_{1^+}(s_{12})\left[-\frac{1}{8}A_1^{0S}(s,s_{13})+
\frac{3}{8}A_0^{1S}(s,s_{13})+
\right.\\

\\
\hskip22mm
+\frac{3}{8}A_0^{0S}(s,s_{13})+\frac{3}{8}A_1^{1S}(s,s_{13})
-\frac{1}{8}A_1^{0S}(s,s_{23})+\frac{3}{8}A_0^{1S}(s,s_{23})+\\

\\
\hskip22mm \left.
+\frac{3}{8}A_0^{0S}(s,s_{23})+\frac{3}{8}A_1^{1S}(s,s_{23})
\right]\\

\\
A_1^{0S}(s,s_{13})=\lambda\, b_{1_S^+}(s_{13})L_{1_S^+}(s_{13})+
K_{1_S^+}(s_{13})\left[\frac{1}{2}A_1^0(s,s_{12})
-\frac{5}{8}A_1^{0S}(s,s_{12})+
\right.\\

\\
\hskip22mm
+\frac{3}{8}A_0^{1S}(s,s_{12})+\frac{3}{8}A_0^{0S}(s,s_{12})
+\frac{3}{8}A_1^{1S}(s,s_{12})+\frac{1}{2}A_1^0(s,s_{23})-\\

\\
\hskip22mm \left.
-\frac{5}{8}A_1^{0S}(s,s_{23})+\frac{3}{8}A_0^{1S}(s,s_{23})+
\frac{3}{8}A_0^{0S}(s,s_{23})+\frac{3}{8}A_1^{1S}(s,s_{23})
\right]\\

\\
A_0^{1S}(s,s_{23})=\lambda\, b_{1_S^-}(s_{23})L_{1_S^-}(s_{23})+
K_{1_S^-}(s_{23})\left[\frac{1}{2}A_1^0(s,s_{12})
-\frac{1}{8}A_1^{0S}(s,s_{12})-
\right.\\

\\
\hskip22mm
-\frac{1}{8}A_0^{1S}(s,s_{12})+\frac{3}{8}A_0^{0S}(s,s_{12})
+\frac{3}{8}A_1^{1S}(s,s_{12})+\frac{1}{2}A_1^0(s,s_{13})-\\

\\
\hskip22mm \left.
-\frac{1}{8}A_1^{0S}(s,s_{13})-\frac{1}{8}A_0^{1S}(s,s_{13})+
\frac{3}{8}A_0^{0S}(s,s_{13})+\frac{3}{8}A_1^{1S}(s,s_{13})
\right]\\

\\
A_0^{0S}(s,s_{13})=\lambda\, b_{0_S^+}(s_{13})L_{0_S^+}(s_{13})+
K_{0_S^+}(s_{13})\left[\frac{1}{2}A_1^0(s,s_{12})
-\frac{1}{8}A_1^{0S}(s,s_{12})+
\right.\\

\\
\hskip22mm
+\frac{3}{8}A_0^{1S}(s,s_{12})-\frac{1}{8}A_0^{0S}(s,s_{12})
+\frac{3}{8}A_1^{1S}(s,s_{12})+\frac{1}{2}A_1^0(s,s_{23})-\\

\\
\hskip22mm \left.
-\frac{1}{8}A_1^{0S}(s,s_{23})+\frac{3}{8}A_0^{1S}(s,s_{23})-
\frac{1}{8}A_0^{0S}(s,s_{23})+\frac{3}{8}A_1^{1S}(s,s_{23})
\right]\\

\\
A_1^{1S}(s,s_{23})=\lambda\, b_{2_S^-}(s_{23})L_{2_S^-}(s_{23})+
K_{2_S^-}(s_{23})\left[\frac{1}{2}A_1^0(s,s_{12})
-\frac{1}{8}A_1^{0S}(s,s_{12})+
\right.\\

\\
\hskip22mm
+\frac{3}{8}A_0^{1S}(s,s_{12})+\frac{3}{8}A_0^{0S}(s,s_{12})
-\frac{1}{8}A_1^{1S}(s,s_{12})+\frac{1}{2}A_1^0(s,s_{13})-\\

\\
\hskip22mm \left.
-\frac{1}{8}A_1^{0S}(s,s_{13})+\frac{3}{8}A_0^{1S}(s,s_{13})+
\frac{3}{8}A_0^{0S}(s,s_{13})-\frac{1}{8}A_1^{1S}(s,s_{13})
\right]\hskip2mm .\\
\end{array}
\right.\eqno (70)$$

\newpage
$\Lambda$ $\frac{3}{2} ^{-}$ $(8,2)$:

$$\left\{
\begin{array}{l}
A_0^1(s,s_{12})=\lambda\, b_{1^-}(s_{12})L_{1^-}(s_{12})+
K_{1^-}(s_{12})\left[\frac{3}{8}A_1^{0S}(s,s_{13})-
\frac{1}{8}A_0^{1S}(s,s_{13})+
\right.\\

\\
\hskip22mm
+\frac{3}{8}A_0^{0S}(s,s_{13})+\frac{3}{8}A_1^{1S}(s,s_{13})
+\frac{3}{8}A_1^{0S}(s,s_{23})-\frac{1}{8}A_0^{1S}(s,s_{23})+\\

\\
\hskip22mm \left.
+\frac{3}{8}A_0^{0S}(s,s_{23})+\frac{3}{8}A_1^{1S}(s,s_{23})
\right]\\

\\
A_1^{0S}(s,s_{13})=\lambda\, b_{1_S^+}(s_{13})L_{1_S^+}(s_{13})+
K_{1_S^+}(s_{13})\left[\frac{1}{2}A_0^1(s,s_{12})
-\frac{1}{8}A_1^{0S}(s,s_{12})-
\right.\\

\\
\hskip22mm
-\frac{1}{8}A_0^{1S}(s,s_{12})+\frac{3}{8}A_0^{0S}(s,s_{12})
+\frac{3}{8}A_1^{1S}(s,s_{12})+\frac{1}{2}A_0^1(s,s_{23})-\\

\\
\hskip22mm \left.
-\frac{1}{8}A_1^{0S}(s,s_{23})-\frac{1}{8}A_0^{1S}(s,s_{23})+
\frac{3}{8}A_0^{0S}(s,s_{23})+\frac{3}{8}A_1^{1S}(s,s_{23})
\right]\\

\\
A_0^{1S}(s,s_{23})=\lambda\, b_{1_S^-}(s_{23})L_{1_S^-}(s_{23})+
K_{1_S^-}(s_{23})\left[\frac{1}{2}A_0^1(s,s_{12})
+\frac{3}{8}A_1^{0S}(s,s_{12})-
\right.\\

\\
\hskip22mm
-\frac{5}{8}A_0^{1S}(s,s_{12})+\frac{3}{8}A_0^{0S}(s,s_{12})
+\frac{3}{8}A_1^{1S}(s,s_{12})+\frac{1}{2}A_0^1(s,s_{13})+\\

\\
\hskip22mm \left.
+\frac{3}{8}A_1^{0S}(s,s_{13})-\frac{5}{8}A_0^{1S}(s,s_{13})+
\frac{3}{8}A_0^{0S}(s,s_{13})+\frac{3}{8}A_1^{1S}(s,s_{13})
\right]\\

\\
A_0^{0S}(s,s_{13})=\lambda\, b_{0_S^+}(s_{13})L_{0_S^+}(s_{13})+
K_{0_S^+}(s_{13})\left[\frac{1}{2}A_0^1(s,s_{12})
+\frac{3}{8}A_1^{0S}(s,s_{12})-
\right.\\

\\
\hskip22mm
-\frac{1}{8}A_0^{1S}(s,s_{12})-\frac{1}{8}A_0^{0S}(s,s_{12})
+\frac{3}{8}A_1^{1S}(s,s_{12})+\frac{1}{2}A_0^1(s,s_{23})+\\

\\
\hskip22mm \left.
+\frac{3}{8}A_1^{0S}(s,s_{23})-\frac{1}{8}A_0^{1S}(s,s_{23})-
\frac{1}{8}A_0^{0S}(s,s_{23})+\frac{3}{8}A_1^{1S}(s,s_{23})
\right]\\

\\
A_1^{1S}(s,s_{23})=\lambda\, b_{2_S^-}(s_{23})L_{2_S^-}(s_{23})+
K_{2_S^-}(s_{23})\left[\frac{1}{2}A_0^1(s,s_{12})
+\frac{3}{8}A_1^{0S}(s,s_{12})-
\right.\\

\\
\hskip22mm
-\frac{1}{8}A_0^{1S}(s,s_{12})+\frac{3}{8}A_0^{0S}(s,s_{12})
-\frac{1}{8}A_1^{1S}(s,s_{12})+\frac{1}{2}A_0^1(s,s_{13})+\\

\\
\hskip22mm \left.
+\frac{3}{8}A_1^{0S}(s,s_{13})-\frac{1}{8}A_0^{1S}(s,s_{13})+
\frac{3}{8}A_0^{0S}(s,s_{13})-\frac{1}{8}A_1^{1S}(s,s_{13})
\right]\hskip2mm .\\
\end{array}
\right.\eqno (71)$$

The system equations of the $\Xi$ $\frac{3}{2} ^{-}$ $(8,2)$ are similar
to the case $\Sigma$ $\frac{3}{2} ^{-}$ $(8,2)$ by replacement
$u\leftrightarrow s$.

As for the case of decuplet, the equations of $\frac{1}{2} ^{-}$ $(8,2)$
multiplet ($l_z=0$) can be determined by the corresponding equation
for $\frac{3}{2} ^{-}$ $(8,2)$ with replacement by the amplitudes:

$1^- \to 0^-$ \, $(A_0^1 \to A_0^{-0})$, \, \, $1^-_S \to 0^-_S$ \,
$(A_0^{1S} \to A_0^{-0S})$,

$2^- \to 1^-$ \, $(A_1^1 \to A_1^{-0})$, \, \, $2^-_S \to 1^-_S$ \,
$(A_1^{1S} \to A_1^{-0S})$.

\newpage
{\bf B3. $(8,4)$ multiplet.}
\vskip 5mm

We consider the states:

$N$ $\frac{5}{2} ^{-}$ $(8,4)$:

$$\left\{
\begin{array}{l}
A_1^0(s,s_{12})=\lambda\, b_{1^+}(s_{12})L_{1^+}(s_{12})+
K_{1^+}(s_{12})\left[\frac{1}{4}A_1^0(s,s_{13})+\frac{3}{4}A_1^1(s,s_{13})+
\right.\\

\\
\hskip22mm \left.
+\frac{1}{4}A_1^0(s,s_{23})+\frac{3}{4}A_1^1(s,s_{23})
\right]\\

\\
A_1^1(s,s_{13})=\lambda\, b_{2^-}(s_{13})L_{2^-}(s_{13})+
K_{2^-}(s_{13})\left[\frac{3}{4}A_1^0(s,s_{12})+\frac{1}{4}A_1^1(s,s_{12})+
\right.\\

\\
\hskip22mm \left.
+\frac{3}{4}A_1^0(s,s_{23})+\frac{1}{4}A_1^1(s,s_{23})
\right] \hskip2mm .\\
\end{array}
\right.\eqno (72)$$

$\Sigma$ $\frac{5}{2} ^{-}$ $(8,4)$:

$$\left\{
\begin{array}{l}
A_1^0(s,s_{12})=\lambda\, b_{1^+}(s_{12})L_{1^+}(s_{12})+
K_{1^+}(s_{12})\left[\frac{1}{4}A_1^{0S}(s,s_{13})+
\frac{3}{4}A_1^{1S}(s,s_{13})+\right.\\

\\
\hskip22mm \left.
+\frac{1}{4}A_1^{0S}(s,s_{23})+\frac{3}{4}A_1^{1S}(s,s_{23})
\right]\\

\\
A_1^{0S}(s,s_{13})=\lambda\, b_{1_S^+}(s_{13})L_{1_S^+}(s_{13})+
K_{1_S^+}(s_{13})\left[\frac{1}{2}A_1^0(s,s_{12})-
\frac{1}{4}A_1^{0S}(s,s_{12})+\right.\\

\\
\hskip22mm \left.
+\frac{3}{4}A_1^{1S}(s,s_{12})+\frac{1}{2}A_1^0(s,s_{23})-
\frac{1}{4}A_1^{0S}(s,s_{23})+\frac{3}{4}A_1^{1S}(s,s_{23})
\right] \\

\\
A_1^{1S}(s,s_{23})=\lambda\, b_{2_S^-}(s_{23})L_{2_S^-}(s_{23})+
K_{2_S^-}(s_{23})\left[\frac{1}{2}A_1^0(s,s_{12})+
\frac{1}{4}A_1^{0S}(s,s_{12})+\right.\\

\\
\hskip22mm \left.
+\frac{1}{4}A_1^{1S}(s,s_{12})+\frac{1}{2}A_1^0(s,s_{13})+
\frac{1}{4}A_1^{0S}(s,s_{13})+\frac{1}{4}A_1^{1S}(s,s_{13})
\right] \hskip2mm .\\
\end{array}
\right.\eqno (73)$$

$\Lambda$ $\frac{5}{2} ^{-}$ $(8,4)$:

$$\left\{
\begin{array}{l}
A_1^1(s,s_{12})=\lambda\, b_{2^-}(s_{12})L_{2^-}(s_{12})+
K_{2^-}(s_{12})\left[\frac{3}{4}A_1^{0S}(s,s_{13})+
\frac{1}{4}A_1^{1S}(s,s_{13})+\right.\\

\\
\hskip22mm \left.
+\frac{3}{4}A_1^{0S}(s,s_{23})+\frac{1}{4}A_1^{1S}(s,s_{23})
\right]\\

\\
A_1^{0S}(s,s_{13})=\lambda\, b_{1_S^+}(s_{13})L_{1_S^+}(s_{13})+
K_{1_S^+}(s_{13})\left[\frac{1}{2}A_1^1(s,s_{12})+
\frac{1}{3}A_1^{0S}(s,s_{12})+\right.\\

\\
\hskip22mm \left.
+\frac{1}{6}A_1^{1S}(s,s_{12})+\frac{1}{2}A_1^1(s,s_{23})+
\frac{1}{3}A_1^{0S}(s,s_{23})+\frac{1}{6}A_1^{1S}(s,s_{23})
\right] \\

\\
A_1^{1S}(s,s_{23})=\lambda\, b_{2_S^-}(s_{23})L_{2_S^-}(s_{23})+
K_{2_S^-}(s_{23})\left[\frac{1}{2}A_1^1(s,s_{12})+
\frac{1}{2}A_1^{0S}(s,s_{12})+\right.\\

\\
\hskip22mm \left.
+\frac{1}{2}A_1^1(s,s_{13})+
\frac{1}{2}A_1^{0S}(s,s_{13})
\right] \hskip2mm .\\
\end{array}
\right.\eqno (74)$$

\newpage
For the particles of $\frac{3}{2} ^{-}$ $(8,4)$ ($l_z=0$) and
$\frac{1}{2} ^{-}$ $(8,4)$ ($l_z=-1$), the equations have the similar
form with replacement by:

$\frac{3}{2} ^{-}$ $(8,4)$: \, \, $2^- \to 1^-$ \,
$(A_1^1 \to A_1^{-0})$, \, \, $2^-_S \to 1^-_S$ \,
$(A_1^{1S} \to A_1^{-0S})$.

$\frac{1}{2} ^{-}$ $(8,4)$: \, \, $2^- \to 0^-$ \,
$(A_1^1 \to A_1^{-1})$, \, \, $2^-_S \to 0^-_S$ \,
$(A_1^{1S} \to A_1^{-1S})$.

\vskip 5mm
{\bf B4. $(1,2)$ multiplet.}
\vskip 5mm

In the case of singlet we obtain only two particles:
$\Lambda$ $\frac{3}{2} ^{-}$ \,($l_z=+1$) and
$\Lambda$ $\frac{1}{2} ^{-}$ \,($l_z=0$).

$\Lambda$ $\frac{3}{2} ^{-}$:

$$\left\{
\begin{array}{l}
A_0^0(s,s_{12})=\lambda\, b_{0^+}(s_{12})L_{0^+}(s_{12})+
K_{0^+}(s_{12})\left[\frac{1}{4}A_0^{0S}(s,s_{13})+
\frac{3}{4}A_1^{1S}(s,s_{13})+\right.\\

\\
\hskip22mm \left.
+\frac{1}{4}A_0^{0S}(s,s_{23})+\frac{3}{4}A_1^{1S}(s,s_{23})
\right]\\

\\
A_0^{0S}(s,s_{13})=\lambda\, b_{0_S^+}(s_{13})L_{0_S^+}(s_{13})+
K_{0_S^+}(s_{13})\left[\frac{1}{2}A_0^0(s,s_{12})-
\frac{1}{4}A_0^{0S}(s,s_{12})+\right.\\

\\
\hskip22mm \left.
+\frac{3}{4}A_1^{1S}(s,s_{12})+\frac{1}{2}A_0^0(s,s_{23})-
\frac{1}{4}A_0^{0S}(s,s_{23})+\frac{3}{4}A_1^{1S}(s,s_{23})
\right] \\

\\
A_1^{1S}(s,s_{23})=\lambda\, b_{2_S^-}(s_{23})L_{2_S^-}(s_{23})+
K_{2_S^-}(s_{23})\left[\frac{1}{2}A_0^0(s,s_{12})+
\frac{1}{4}A_0^{0S}(s,s_{12})+\right.\\

\\
\hskip22mm \left.
+\frac{1}{4}A_1^{1S}(s,s_{12})+\frac{1}{2}A_0^0(s,s_{13})+
\frac{1}{4}A_0^{0S}(s,s_{13})+\frac{1}{4}A_1^{1S}(s,s_{13})
\right] \hskip2mm .\\
\end{array}
\right.\eqno (75)$$

For the $\Lambda$ $\frac{1}{2} ^{-}$ we replace by
$2^-_S \to 1^-_S$ \, $(A_1^{1S} \to A_1^{-0S})$ and obtain:

$$\left\{
\begin{array}{l}
A_0^0(s,s_{12})=\lambda\, b_{0^+}(s_{12})L_{0^+}(s_{12})+
K_{0^+}(s_{12})\left[\frac{1}{4}A_0^{0S}(s,s_{13})+
\frac{3}{4}A_1^{-0S}(s,s_{13})+\right.\\

\\
\hskip22mm \left.
+\frac{1}{4}A_0^{0S}(s,s_{23})+\frac{3}{4}A_1^{-0S}(s,s_{23})
\right]\\

\\
A_0^{0S}(s,s_{13})=\lambda\, b_{0_S^+}(s_{13})L_{0_S^+}(s_{13})+
K_{0_S^+}(s_{13})\left[\frac{1}{2}A_0^0(s,s_{12})-
\frac{1}{4}A_0^{0S}(s,s_{12})+\right.\\

\\
\hskip22mm \left.
+\frac{3}{4}A_1^{-0S}(s,s_{12})+\frac{1}{2}A_0^0(s,s_{23})-
\frac{1}{4}A_0^{0S}(s,s_{23})+\frac{3}{4}A_1^{-0S}(s,s_{23})
\right] \\

\\
A_1^{-0S}(s,s_{23})=\lambda\, b_{1_S^-}(s_{23})L_{1_S^-}(s_{23})+
K_{1_S^-}(s_{23})\left[\frac{1}{2}A_0^0(s,s_{12})+
\frac{1}{4}A_0^{0S}(s,s_{12})+\right.\\

\\
\hskip22mm \left.
+\frac{1}{4}A_1^{-0S}(s,s_{12})+\frac{1}{2}A_0^0(s,s_{13})+
\frac{1}{4}A_0^{0S}(s,s_{13})+\frac{1}{4}A_1^{-0S}(s,s_{13})
\right] \hskip2mm .\\
\end{array}
\right.\eqno (76)$$

\newpage
{\bf Appendix C. The system equations of reduced amplitude of the
multiplets ${\bf (70,1^-)}$.}
\vskip 5mm
{\bf C1. The equations of ${\bf (10,2)}$ multiplet.}
\vskip 5mm

$\Delta$ $\frac{3}{2} ^{-}$ $(10,2)$:

$$\left\{
\begin{array}{l}
\alpha_1^0(s,s_0)=\lambda+\frac{1}{2}\hskip2mm\alpha_1^0(s,s_0)
\hskip2mm I_{1^+ 1^+}(s,s_0)\hskip40mm 1^+\\
\hskip25mm
+\frac{3}{2}\hskip2mm\alpha_0^1(s,s_0)\hskip2mm I_{1^+ 1^-}(s,s_0)\hskip2mm
\frac{b_{1^-}(s_0)}{b_{1^+}(s_0)}\\
\alpha_0^1(s,s_0)=\lambda+\frac{3}{2}\hskip2mm\alpha_1^0(s,s_0)
\hskip2mm I_{1^- 1^+}(s,s_0)
\hskip2mm\frac{b_{1^+}(s_0)}{b_{1^-}(s_0)}\hskip27mm 1^-\\
\hskip25mm
+\frac{1}{2}\hskip2mm\alpha_0^1(s,s_0)\hskip2mm I_{1^- 1^-}(s,s_0)
\hskip2mm .\\
\end{array} \right.\eqno (77)$$

Here $\alpha_1^0$ and $\alpha_0^1$ are the reduced diquark amplitudes
with $J^p=1^+$ and $1^-$, in the case of $\Delta$ $\frac{3}{2} ^{-}$
$(10,2)$ one obtain only these two amplitudes.

The equations for $\Sigma$ $\frac{3}{2} ^{-}$ $(10,2)$:

$$\left\{
\begin{array}{l}
\alpha_1^0(s,s_0)=\lambda+\frac{1}{2}\hskip2mm\alpha_1^{0S}(s,s_0)
\hskip2mm I_{1^+ 1^+_S}(s,s_0)
\hskip2mm\frac{b_{1^+_s}(s_0)}{b_{1^+}(s_0)}\hskip40mm 1^+\\
\hskip25mm
+\frac{3}{2}\hskip2mm\alpha_0^{1S}(s,s_0)\hskip2mm I_{1^+ 1^-_S}(s,s_0)
\hskip2mm\frac{b_{1^-_s}(s_0)}{b_{1^+}(s_0)}\\
\alpha_1^{0S}(s,s_0)=\lambda
+\alpha_1^0(s,s_0)\hskip2mm I_{1^+_S 1^+}(s,s_0)
\hskip2mm\frac{b_{1^+}(s_0)}{b_{1^+_s}(s_0)}\hskip45mm 1^+_S\\
\hskip25mm
-\frac{1}{2}\hskip2mm\alpha_1^{0S}(s,s_0)
\hskip2mm I_{1^+_S 1^+_S}(s,s_0)
+\frac{3}{2}\hskip2mm\alpha_0^{1S}(s,s_0)\hskip2mm I_{1^+_S 1^-_S}(s,s_0)
\hskip2mm\frac{b_{1^-_s}(s_0)}{b_{1^+_s}(s_0)}\\
\alpha_0^{1S}(s,s_0)=\lambda+\alpha_1^0(s,s_0)\hskip2mm I_{1^-_S 1^+}(s,s_0)
\hskip2mm\frac{b_{1^+}(s_0)}{b_{1^-_s}(s_0)}\hskip45mm 1^-_S\\
\hskip25mm
+\frac{1}{2}\hskip2mm\alpha_1^{0S}(s,s_0)
\hskip2mm I_{1^-_S 1^+_S}(s,s_0)
\hskip2mm\frac{b_{1^+_s}(s_0)}{b_{1^-_s}(s_0)}
+\frac{1}{2}\hskip2mm\alpha_0^{1S}(s,s_0)\hskip2mm I_{1^-_S 1^-_S}(s,s_0)
\hskip2mm .\\
\end{array} \right.\eqno (78)$$

Here we use the reduced amplitudes of three diquarks:
$1^+$, $1^+_s$, $1^-_s$.

The $\Xi$ system equations is similar to the case $\Sigma$ with the
replacement by $u\leftrightarrow s$:

$$\left\{
\begin{array}{l}
\alpha_1^{0SS}(s,s_0)=\lambda+\frac{1}{2}\hskip2mm\alpha_1^{0S}(s,s_0)
\hskip2mm I_{1^+_{SS} 1^+_S}(s,s_0)
\hskip2mm\frac{b_{1^+_s}(s_0)}{b_{1^+_{ss}}(s_0)}\hskip40mm 1^+_{SS}\\
\hskip25mm
+\frac{3}{2}\hskip2mm\alpha_0^{1S}(s,s_0)\hskip2mm I_{1^+_{SS} 1^-_S}(s,s_0)
\hskip2mm\frac{b_{1^-_s}(s_0)}{b_{1^+_{ss}}(s_0)}\\
\alpha_1^{0S}(s,s_0)=\lambda
+\alpha_1^{0SS}(s,s_0)\hskip2mm I_{1^+_S 1^+_{SS}}(s,s_0)
\hskip2mm\frac{b_{1^+_{ss}}(s_0)}{b_{1^+_s}(s_0)}\hskip45mm 1^+_S\\
\hskip25mm
-\frac{1}{2}\hskip2mm\alpha_1^{0S}(s,s_0)
\hskip2mm I_{1^+_S 1^+_S}(s,s_0)
+\frac{3}{2}\hskip2mm\alpha_0^{1S}(s,s_0)\hskip2mm I_{1^+_S 1^-_S}(s,s_0)
\hskip2mm\frac{b_{1^-_s}(s_0)}{b_{1^+_s}(s_0)}\\
\alpha_0^{1S}(s,s_0)=\lambda+\alpha_1^{0SS}(s,s_0)\hskip2mm
I_{1^-_S 1^+_{SS}}(s,s_0)
\hskip2mm\frac{b_{1^+_{ss}}(s_0)}{b_{1^-_s}(s_0)}\hskip45mm 1^-_S\\
\hskip25mm
+\frac{1}{2}\hskip2mm\alpha_1^{0S}(s,s_0)
\hskip2mm I_{1^-_S 1^+_S}(s,s_0)
\hskip2mm\frac{b_{1^+_s}(s_0)}{b_{1^-_s}(s_0)}
+\frac{1}{2}\hskip2mm\alpha_0^{1S}(s,s_0)\hskip2mm I_{1^-_S 1^-_S}(s,s_0)
\hskip2mm .\\
\end{array} \right.\eqno (79)$$

By analogy with the $\Delta$ equations we calculate the $\Omega$ equations
with replacement by $u\rightarrow s$.

Then one considers the multiplet $\frac{1}{2}^-$ $(10,2)$. In this case
the $l_z$ equal to $0$ ($l_z=0$). The equations of $\frac{1}{2}^-$
multiplets are similar to the $\frac{3}{2}^-$ with replacement by reduced
amplitudes: $1^-\rightarrow 0^-$, $\alpha_0^1\rightarrow \alpha_0^{-0}$
and $1^-_S\rightarrow 0^-_S$, $\alpha_0^{1S}\rightarrow \alpha_0^{-0S}$.

\newpage
The multiplet $(10,2)$:

$\Delta$ $\frac{1}{2} ^{-}$ $(10,2)$:

$$\left\{
\begin{array}{l}
\alpha_1^0(s,s_0)=\lambda+\frac{1}{2}\hskip2mm\alpha_1^0(s,s_0)
\hskip2mm I_{1^+ 1^+}(s,s_0)\hskip43mm 1^+\\
\hskip25mm
+\frac{3}{2}\hskip2mm\alpha_0^{-0}(s,s_0)\hskip2mm I_{1^+ 0^-}(s,s_0)
\hskip2mm\frac{b_{0^-}(s_0)}{b_{1^+}(s_0)}\\
\alpha_0^{-0}(s,s_0)=\lambda+\frac{3}{2}\hskip2mm\alpha_1^0(s,s_0)
\hskip2mm I_{0^- 1^+}(s,s_0)
\hskip2mm\frac{b_{1^+}(s_0)}{b_{0^-}(s_0)}\hskip28mm 0^-\\
\hskip25mm
+\frac{1}{2}\hskip2mm\alpha_0^{-0}(s,s_0)\hskip2mm I_{0^- 0^-}(s,s_0)
\hskip2mm .\\
\end{array} \right.\eqno (80)$$

$\Sigma$ $\frac{1}{2} ^{-}$ $(10,2)$:

$$\left\{
\begin{array}{l}
\alpha_1^0(s,s_0)=\lambda+\frac{1}{2}\hskip2mm\alpha_1^{0S}(s,s_0)
\hskip2mm I_{1^+ 1^+_S}(s,s_0)
\hskip2mm\frac{b_{1^+_s}(s_0)}{b_{1^+}(s_0)}\hskip40mm 1^+\\
\hskip25mm
+\frac{3}{2}\hskip2mm\alpha_0^{-0S}(s,s_0)\hskip2mm I_{1^+ 0^-_S}(s,s_0)
\hskip2mm\frac{b_{0^-_s}(s_0)}{b_{1^+}(s_0)}\\
\alpha_1^{0S}(s,s_0)=\lambda
+\alpha_1^0(s,s_0)\hskip2mm I_{1^+_S 1^+}(s,s_0)
\hskip2mm\frac{b_{1^+}(s_0)}{b_{1^+_s}(s_0)}\hskip45mm 1^+_S\\
\hskip25mm
-\frac{1}{2}\hskip2mm\alpha_1^{0S}(s,s_0)
\hskip2mm I_{1^+_S 1^+_S}(s,s_0)
+\frac{3}{2}\hskip2mm\alpha_0^{-0S}(s,s_0)\hskip2mm I_{1^+_S 0^-_S}(s,s_0)
\hskip2mm\frac{b_{0^-_s}(s_0)}{b_{1^+_s}(s_0)}\\
\alpha_0^{-0S}(s,s_0)=\lambda+\alpha_1^0(s,s_0)
\hskip2mm I_{0^-_S 1^+}(s,s_0)
\hskip2mm\frac{b_{1^+}(s_0)}{b_{0^-_s}(s_0)}\hskip43mm 0^-_S\\
\hskip25mm
+\frac{1}{2}\hskip2mm\alpha_1^{0S}(s,s_0)
\hskip2mm I_{0^-_S 1^+_S}(s,s_0)
\hskip2mm\frac{b_{1^+_s}(s_0)}{b_{0^-_s}(s_0)}
+\frac{1}{2}\hskip2mm\alpha_0^{-0S}(s,s_0)\hskip2mm I_{0^-_S 0^-_S}(s,s_0)
\hskip2mm .\\
\end{array} \right.\eqno (81)$$

We can obtain that the equations of $\Sigma$ are similar to $\Delta$ with
the replacement by $s\rightarrow u$.

\vskip 5mm
{\bf C2. The equations of ${\bf (8,2)}$ multiplet.}
\vskip 5mm

$N$ $\frac{3}{2} ^{-}$ $(8,2)$:

$$\left\{
\begin{array}{l}
\alpha_1^0(s,s_0)=\lambda-\frac{1}{4}\hskip2mm\alpha_1^0(s,s_0)
\hskip2mm I_{1^+ 1^+}(s,s_0)
+\frac{3}{4}\hskip2mm\alpha_0^1(s,s_0)\hskip2mm I_{1^+ 1^-}(s,s_0)\hskip2mm
\frac{b_{1^-}(s_0)}{b_{1^+}(s_0)}\hskip4mm 1^+\\

\\
\hskip14mm
+\frac{3}{4}\hskip2mm\alpha_0^0(s,s_0)\hskip2mm I_{1^+ 0^+}(s,s_0)\hskip2mm
\frac{b_{0^+}(s_0)}{b_{1^+}(s_0)}
+\frac{3}{4}\hskip2mm\alpha_1^1(s,s_0)\hskip2mm I_{1^+ 2^-}(s,s_0)\hskip2mm
\frac{b_{2^-}(s_0)}{b_{1^+}(s_0)}\\

\\
\alpha_0^1(s,s_0)=\lambda+\frac{3}{4}\hskip2mm\alpha_1^0(s,s_0)
\hskip2mm I_{1^- 1^+}(s,s_0)\hskip2mm
\frac{b_{1^+}(s_0)}{b_{1^-}(s_0)}
-\frac{1}{4}\hskip2mm\alpha_0^1(s,s_0)\hskip2mm I_{1^- 1^-}(s,s_0)
\hskip5mm 1^-\\

\\
\hskip14mm
+\frac{3}{4}\hskip2mm\alpha_0^0(s,s_0)\hskip2mm I_{1^- 0^+}(s,s_0)\hskip2mm
\frac{b_{0^+}(s_0)}{b_{1^-}(s_0)}
+\frac{3}{4}\hskip2mm\alpha_1^1(s,s_0)\hskip2mm I_{1^- 2^-}(s,s_0)\hskip2mm
\frac{b_{2^-}(s_0)}{b_{1^-}(s_0)}\\

\\
\alpha_0^0(s,s_0)=\lambda+\frac{3}{4}\hskip2mm\alpha_1^0(s,s_0)
\hskip2mm I_{0^+ 1^+}(s,s_0)\hskip2mm\frac{b_{1^+}(s_0)}{b_{0^+}(s_0)}
+\frac{3}{4}\hskip2mm\alpha_0^1(s,s_0)\hskip2mm I_{0^+ 1^-}(s,s_0)\cdot
\hskip2mm 0^+\\

\\
\hskip14mm
\cdot\frac{b_{1^-}(s_0)}{b_{0^+}(s_0)}-\frac{1}{4}
\hskip2mm\alpha_0^0(s,s_0)\hskip2mm I_{0^+ 0^+}(s,s_0)
+\frac{3}{4}\hskip2mm\alpha_1^1(s,s_0)\hskip2mm I_{0^+ 2^-}(s,s_0)\hskip2mm
\frac{b_{2^-}(s_0)}{b_{0^+}(s_0)}\\

\\
\alpha_1^1(s,s_0)=\lambda+\frac{3}{4}\hskip2mm\alpha_1^0(s,s_0)
\hskip2mm I_{2^- 1^+}(s,s_0)\hskip2mm\frac{b_{1^+}(s_0)}{b_{2^-}(s_0)}
+\frac{3}{4}\hskip2mm\alpha_0^1(s,s_0)\hskip2mm I_{2^- 1^-}(s,s_0)\cdot
\hskip2mm 2^-\\

\\
\hskip14mm
\cdot\frac{b_{1^-}(s_0)}{b_{2^-}(s_0)}+\frac{3}{4}
\hskip2mm\alpha_0^0(s,s_0)\hskip2mm I_{2^- 0^+}(s,s_0)
\hskip2mm\frac{b_{0^+}(s_0)}{b_{2^-}(s_0)}
-\frac{1}{4}\hskip2mm\alpha_1^1(s,s_0)\hskip2mm I_{2^- 2^-}(s,s_0)
\hskip2mm .\\
\end{array} \right.\eqno (82)$$

\newpage
$\Sigma$ $\frac{3}{2} ^{-}$ $(8,2)$:

$$\left\{
\begin{array}{l}
\alpha_1^0(s,s_0)=\lambda-\frac{1}{4}\hskip2mm\alpha_1^{0S}(s,s_0)
\hskip2mm I_{1^+ 1^+_S}(s,s_0)\hskip2mm\frac{b_{1^+_s}(s_0)}{b_{1^+}(s_0)}
\hskip40mm 1^+\\

\\
\hskip14mm
+\frac{3}{4}\hskip2mm\alpha_0^{1S}(s,s_0)\hskip2mm I_{1^+ 1^-_S}(s,s_0)
\hskip2mm\frac{b_{1^-_s}(s_0)}{b_{1^+}(s_0)}
+\frac{3}{4}\hskip2mm\alpha_0^{0S}(s,s_0)\hskip2mm I_{1^+ 0^+_S}(s,s_0)
\hskip2mm\frac{b_{0^+_s}(s_0)}{b_{1^+}(s_0)}\\

\\
\hskip14mm
+\frac{3}{4}\hskip2mm\alpha_1^{1S}(s,s_0)\hskip2mm I_{1^+ 2^-_S}(s,s_0)
\hskip2mm\frac{b_{2^-_s}(s_0)}{b_{1^+}(s_0)}\\

\\
\alpha_1^{0S}(s,s_0)=\lambda+\alpha_1^0(s,s_0)
\hskip2mm I_{1^+_S 1^+}(s,s_0)\hskip2mm\frac{b_{1^+}(s_0)}{b_{1^+_s}(s_0)}
\hskip45mm 1^+_s\\

\\
\hskip14mm
-\frac{5}{4}\hskip2mm\alpha_1^{0S}(s,s_0)
\hskip2mm I_{1^+_S 1^+_S}(s,s_0)
+\frac{3}{4}\hskip2mm\alpha_0^{1S}(s,s_0)\hskip2mm I_{1^+_S 1^-_S}(s,s_0)
\hskip2mm\frac{b_{1^-_s}(s_0)}{b_{1^+_s}(s_0)}\\

\\
\hskip14mm
+\frac{3}{4}\hskip2mm\alpha_0^{0S}(s,s_0)\hskip2mm I_{1^+_S 0^+_S}(s,s_0)
\hskip2mm\frac{b_{0^+_s}(s_0)}{b_{1^+_s}(s_0)}
+\frac{3}{4}\hskip2mm\alpha_1^{1S}(s,s_0)\hskip2mm I_{1^+_S 2^-_S}(s,s_0)
\hskip2mm\frac{b_{2^-_s}(s_0)}{b_{1^+_s}(s_0)}\\

\\
\alpha_0^{1S}(s,s_0)=\lambda+\alpha_1^0(s,s_0)
\hskip2mm I_{1^-_S 1^+}(s,s_0)\hskip2mm\frac{b_{1^+}(s_0)}{b_{1^-_s}(s_0)}
\hskip45mm 1^-_s\\

\\
\hskip14mm
-\frac{1}{4}\hskip2mm\alpha_1^{0S}(s,s_0)
\hskip2mm I_{1^-_S 1^+_S}(s,s_0)
\hskip2mm\frac{b_{1^+_s}(s_0)}{b_{1^-_s}(s_0)}
-\frac{1}{4}\hskip2mm\alpha_0^{1S}(s,s_0)\hskip2mm I_{1^-_S 1^-_S}(s,s_0)\\

\\
\hskip14mm
+\frac{3}{4}\hskip2mm\alpha_0^{0S}(s,s_0)\hskip2mm I_{1^-_S 0^+_S}(s,s_0)
\hskip2mm\frac{b_{0^+_s}(s_0)}{b_{1^-_s}(s_0)}
+\frac{3}{4}\hskip2mm\alpha_1^{1S}(s,s_0)\hskip2mm I_{1^-_S 2^-_S}(s,s_0)
\hskip2mm\frac{b_{2^-_s}(s_0)}{b_{1^-_s}(s_0)}\\

\\
\alpha_0^{0S}(s,s_0)=\lambda+\alpha_1^0(s,s_0)
\hskip2mm I_{0^+_S 1^+}(s,s_0)\hskip2mm\frac{b_{1^+}(s_0)}{b_{0^+_s}(s_0)}
\hskip45mm 0^+_s\\

\\
\hskip14mm
-\frac{1}{4}\hskip2mm\alpha_1^{0S}(s,s_0)
\hskip2mm I_{0^+_S 1^+_S}(s,s_0)
\hskip2mm\frac{b_{1^+_s}(s_0)}{b_{0^+_s}(s_0)}
+\frac{3}{4}\hskip2mm\alpha_0^{1S}(s,s_0)\hskip2mm I_{0^+_S 1^-_S}(s,s_0)
\hskip2mm\frac{b_{1^-_s}(s_0)}{b_{0^+_s}(s_0)}\\

\\
\hskip14mm
-\frac{1}{4}\hskip2mm\alpha_0^{0S}(s,s_0)\hskip2mm I_{0^+_S 0^+_S}(s,s_0)
+\frac{3}{4}\hskip2mm\alpha_1^{1S}(s,s_0)\hskip2mm I_{0^+_S 2^-_S}(s,s_0)
\hskip2mm\frac{b_{2^-_s}(s_0)}{b_{0^+_s}(s_0)}\\

\\
\alpha_1^{1S}(s,s_0)=\lambda+\alpha_1^0(s,s_0)
\hskip2mm I_{2^-_S 1^+}(s,s_0)\hskip2mm\frac{b_{1^+}(s_0)}{b_{2^-_s}(s_0)}
\hskip45mm 2^-_s\\

\\
\hskip14mm
-\frac{1}{4}\hskip2mm\alpha_1^{0S}(s,s_0)
\hskip2mm I_{2^-_S 1^+_S}(s,s_0)
\hskip2mm\frac{b_{1^+_s}(s_0)}{b_{2^-_s}(s_0)}
+\frac{3}{4}\hskip2mm\alpha_0^{1S}(s,s_0)\hskip2mm I_{2^-_S 1^-_S}(s,s_0)
\hskip2mm\frac{b_{1^-_s}(s_0)}{b_{2^-_s}(s_0)}\\

\\
\hskip14mm
+\frac{3}{4}\hskip2mm\alpha_0^{0S}(s,s_0)\hskip2mm I_{2^-_S 0^+_S}(s,s_0)
\hskip2mm\frac{b_{0^+_s}(s_0)}{b_{2^-_s}(s_0)}
-\frac{1}{4}\hskip2mm\alpha_1^{1S}(s,s_0)\hskip2mm I_{2^-_S 2^-_S}(s,s_0)
\hskip2mm .\\
\end{array} \right.\eqno (83)$$

\newpage
By analogy with the case $(10,2)$, the equations for the
$\Xi$ $\frac{3}{2} ^{-}$ $(8,2)$ are similar to $\Sigma$ $\frac{3}{2} ^{-}$
with replacement by $u\leftrightarrow s$:

$$\left\{
\begin{array}{l}
\alpha_1^{0SS}(s,s_0)=\lambda-\frac{1}{4}\hskip2mm\alpha_1^{0S}(s,s_0)
\hskip2mm I_{1^+_{SS} 1^+_S}(s,s_0)\hskip2mm
\frac{b_{1^+_s}(s_0)}{b_{1^+_{ss}}(s_0)}
\hskip40mm 1^+_{ss}\\

\\
\hskip14mm
+\frac{3}{4}\hskip2mm\alpha_0^{1S}(s,s_0)\hskip2mm I_{1^+_{SS} 1^-_S}(s,s_0)
\hskip2mm\frac{b_{1^-_s}(s_0)}{b_{1^+_{ss}}(s_0)}
+\frac{3}{4}\hskip2mm\alpha_0^{0S}(s,s_0)\hskip2mm I_{1^+_{SS} 0^+_S}(s,s_0)
\hskip2mm\frac{b_{0^+_s}(s_0)}{b_{1^+_{ss}}(s_0)}\\

\\
\hskip14mm
+\frac{3}{4}\hskip2mm\alpha_1^{1S}(s,s_0)\hskip2mm I_{1^+_{SS} 2^-_S}(s,s_0)
\hskip2mm\frac{b_{2^-_s}(s_0)}{b_{1^+_{ss}}(s_0)}\\

\\
\alpha_1^{0S}(s,s_0)=\lambda+\alpha_1^{0SS}(s,s_0)
\hskip2mm I_{1^+_S 1^+_{SS}}(s,s_0)\hskip2mm
\frac{b_{1^+_{ss}}(s_0)}{b_{1^+_s}(s_0)}
\hskip45mm 1^+_s\\

\\
\hskip14mm
-\frac{5}{4}\hskip2mm\alpha_1^{0S}(s,s_0)
\hskip2mm I_{1^+_S 1^+_S}(s,s_0)
+\frac{3}{4}\hskip2mm\alpha_0^{1S}(s,s_0)\hskip2mm I_{1^+_S 1^-_S}(s,s_0)
\hskip2mm\frac{b_{1^-_s}(s_0)}{b_{1^+_s}(s_0)}\\

\\
\hskip14mm
+\frac{3}{4}\hskip2mm\alpha_0^{0S}(s,s_0)\hskip2mm I_{1^+_S 0^+_S}(s,s_0)
\hskip2mm\frac{b_{0^+_s}(s_0)}{b_{1^+_s}(s_0)}
+\frac{3}{4}\hskip2mm\alpha_1^{1S}(s,s_0)\hskip2mm I_{1^+_S 2^-_S}(s,s_0)
\hskip2mm\frac{b_{2^-_s}(s_0)}{b_{1^+_s}(s_0)}\\

\\
\alpha_0^{1S}(s,s_0)=\lambda+\alpha_1^{0SS}(s,s_0)
\hskip2mm I_{1^-_S 1^+_{SS}}(s,s_0)\hskip2mm\frac{b_{1^+_{ss}}(s_0)}
{b_{1^-_s}(s_0)}
\hskip45mm 1^-_s\\

\\
\hskip14mm
-\frac{1}{4}\hskip2mm\alpha_1^{0S}(s,s_0)
\hskip2mm I_{1^-_S 1^+_S}(s,s_0)
\hskip2mm\frac{b_{1^+_s}(s_0)}{b_{1^-_s}(s_0)}
-\frac{1}{4}\hskip2mm\alpha_0^{1S}(s,s_0)\hskip2mm I_{1^-_S 1^-_S}(s,s_0)\\

\\
\hskip14mm
+\frac{3}{4}\hskip2mm\alpha_0^{0S}(s,s_0)\hskip2mm I_{1^-_S 0^+_S}(s,s_0)
\hskip2mm\frac{b_{0^+_s}(s_0)}{b_{1^-_s}(s_0)}
+\frac{3}{4}\hskip2mm\alpha_1^{1S}(s,s_0)\hskip2mm I_{1^-_S 2^-_S}(s,s_0)
\hskip2mm\frac{b_{2^-_s}(s_0)}{b_{1^-_s}(s_0)}\\

\\
\alpha_0^{0S}(s,s_0)=\lambda+\alpha_1^{0SS}(s,s_0)
\hskip2mm I_{0^+_S 1^+_{SS}}(s,s_0)\hskip2mm\frac{b_{1^+_{ss}}(s_0)}
{b_{0^+_s}(s_0)}
\hskip45mm 0^+_s\\

\\
\hskip14mm
-\frac{1}{4}\hskip2mm\alpha_1^{0S}(s,s_0)
\hskip2mm I_{0^+_S 1^+_S}(s,s_0)
\hskip2mm\frac{b_{1^+_s}(s_0)}{b_{0^+_s}(s_0)}
+\frac{3}{4}\hskip2mm\alpha_0^{1S}(s,s_0)\hskip2mm I_{0^+_S 1^-_S}(s,s_0)
\hskip2mm\frac{b_{1^-_s}(s_0)}{b_{0^+_s}(s_0)}\\

\\
\hskip14mm
-\frac{1}{4}\hskip2mm\alpha_0^{0S}(s,s_0)\hskip2mm I_{0^+_S 0^+_S}(s,s_0)
+\frac{3}{4}\hskip2mm\alpha_1^{1S}(s,s_0)\hskip2mm I_{0^+_S 2^-_S}(s,s_0)
\hskip2mm\frac{b_{2^-_s}(s_0)}{b_{0^+_s}(s_0)}\\

\\
\alpha_1^{1S}(s,s_0)=\lambda+\alpha_1^{0SS}(s,s_0)
\hskip2mm I_{2^-_S 1^+_{SS}}(s,s_0)\hskip2mm\frac{b_{1^+_{ss}}(s_0)}
{b_{2^-_s}(s_0)}
\hskip45mm 2^-_s\\

\\
\hskip14mm
-\frac{1}{4}\hskip2mm\alpha_1^{0S}(s,s_0)
\hskip2mm I_{2^-_S 1^+_S}(s,s_0)
\hskip2mm\frac{b_{1^+_s}(s_0)}{b_{2^-_s}(s_0)}
+\frac{3}{4}\hskip2mm\alpha_0^{1S}(s,s_0)\hskip2mm I_{2^-_S 1^-_S}(s,s_0)
\hskip2mm\frac{b_{1^-_s}(s_0)}{b_{2^-_s}(s_0)}\\

\\
\hskip14mm
+\frac{3}{4}\hskip2mm\alpha_0^{0S}(s,s_0)\hskip2mm I_{2^-_S 0^+_S}(s,s_0)
\hskip2mm\frac{b_{0^+_s}(s_0)}{b_{2^-_s}(s_0)}
-\frac{1}{4}\hskip2mm\alpha_1^{1S}(s,s_0)\hskip2mm I_{2^-_S 2^-_S}(s,s_0)
\hskip2mm .\\
\end{array} \right.\eqno (84)$$

\newpage
$\Lambda$ $\frac{3}{2} ^{-}$ $(8,2)$:

$$\left\{
\begin{array}{l}
\alpha_1^0(s,s_0)=\lambda+\frac{3}{4}\hskip2mm\alpha_1^{0S}(s,s_0)
\hskip2mm I_{1^+ 1^+_S}(s,s_0)\hskip2mm\frac{b_{1^+_s}(s_0)}{b_{1^+}(s_0)}
\hskip40mm 1^+\\

\\
\hskip14mm
-\frac{1}{4}\hskip2mm\alpha_0^{1S}(s,s_0)\hskip2mm I_{1^+ 1^-_S}(s,s_0)
\hskip2mm\frac{b_{1^-_s}(s_0)}{b_{1^+}(s_0)}
+\frac{3}{4}\hskip2mm\alpha_0^{0S}(s,s_0)\hskip2mm I_{1^+ 0^+_S}(s,s_0)
\hskip2mm\frac{b_{0^+_s}(s_0)}{b_{1^+}(s_0)}\\

\\
\hskip14mm
+\frac{3}{4}\hskip2mm\alpha_1^{1S}(s,s_0)\hskip2mm I_{1^+ 2^-_S}(s,s_0)
\hskip2mm\frac{b_{2^-_s}(s_0)}{b_{1^+}(s_0)}\\

\\
\alpha_1^{0S}(s,s_0)=\lambda+\alpha_1^0(s,s_0)
\hskip2mm I_{1^+_S 1^+}(s,s_0)\hskip2mm\frac{b_{1^+}(s_0)}{b_{1^+_s}(s_0)}
\hskip45mm 1^+_s\\

\\
\hskip14mm
-\frac{1}{4}\hskip2mm\alpha_1^{0S}(s,s_0)
\hskip2mm I_{1^+_S 1^+_S}(s,s_0)
-\frac{1}{4}\hskip2mm\alpha_0^{1S}(s,s_0)\hskip2mm I_{1^+_S 1^-_S}(s,s_0)
\hskip2mm\frac{b_{1^-_s}(s_0)}{b_{1^+_s}(s_0)}\\

\\
\hskip14mm
+\frac{3}{4}\hskip2mm\alpha_0^{0S}(s,s_0)\hskip2mm I_{1^+_S 0^+_S}(s,s_0)
\hskip2mm\frac{b_{0^+_s}(s_0)}{b_{1^+_s}(s_0)}
+\frac{3}{4}\hskip2mm\alpha_1^{1S}(s,s_0)\hskip2mm I_{1^+_S 2^-_S}(s,s_0)
\hskip2mm\frac{b_{2^-_s}(s_0)}{b_{1^+_s}(s_0)}\\

\\
\alpha_0^{1S}(s,s_0)=\lambda+\alpha_1^0(s,s_0)
\hskip2mm I_{1^-_S 1^+}(s,s_0)\hskip2mm\frac{b_{1^+}(s_0)}{b_{1^-_s}(s_0)}
\hskip45mm 1^-_s\\

\\
\hskip14mm
+\frac{3}{4}\hskip2mm\alpha_1^{0S}(s,s_0)
\hskip2mm I_{1^-_S 1^+_S}(s,s_0)
\hskip2mm\frac{b_{1^+_s}(s_0)}{b_{1^-_s}(s_0)}
-\frac{5}{4}\hskip2mm\alpha_0^{1S}(s,s_0)\hskip2mm I_{1^-_S 1^-_S}(s,s_0)\\

\\
\hskip14mm
+\frac{3}{4}\hskip2mm\alpha_0^{0S}(s,s_0)\hskip2mm I_{1^-_S 0^+_S}(s,s_0)
\hskip2mm\frac{b_{0^+_s}(s_0)}{b_{1^-_s}(s_0)}
+\frac{3}{4}\hskip2mm\alpha_1^{1S}(s,s_0)\hskip2mm I_{1^-_S 2^-_S}(s,s_0)
\hskip2mm\frac{b_{2^-_s}(s_0)}{b_{1^-_s}(s_0)}\\

\\
\alpha_0^{0S}(s,s_0)=\lambda+\alpha_1^0(s,s_0)
\hskip2mm I_{0^+_S 1^+}(s,s_0)\hskip2mm\frac{b_{1^+}(s_0)}{b_{0^+_s}(s_0)}
\hskip45mm 0^+_s\\

\\
\hskip14mm
+\frac{3}{4}\hskip2mm\alpha_1^{0S}(s,s_0)
\hskip2mm I_{0^+_S 1^+_S}(s,s_0)
\hskip2mm\frac{b_{1^+_s}(s_0)}{b_{0^+_s}(s_0)}
-\frac{1}{4}\hskip2mm\alpha_0^{1S}(s,s_0)\hskip2mm I_{0^+_S 1^-_S}(s,s_0)
\hskip2mm\frac{b_{1^-_s}(s_0)}{b_{0^+_s}(s_0)}\\

\\
\hskip14mm
-\frac{1}{4}\hskip2mm\alpha_0^{0S}(s,s_0)\hskip2mm I_{0^+_S 0^+_S}(s,s_0)
+\frac{3}{4}\hskip2mm\alpha_1^{1S}(s,s_0)\hskip2mm I_{0^+_S 2^-_S}(s,s_0)
\hskip2mm\frac{b_{2^-_s}(s_0)}{b_{0^+_s}(s_0)}\\

\\
\alpha_1^{1S}(s,s_0)=\lambda+\alpha_1^0(s,s_0)
\hskip2mm I_{2^-_S 1^+}(s,s_0)\hskip2mm\frac{b_{1^+}(s_0)}{b_{2^-_s}(s_0)}
\hskip45mm 2^-_s\\

\\
\hskip14mm
+\frac{3}{4}\hskip2mm\alpha_1^{0S}(s,s_0)
\hskip2mm I_{2^-_S 1^+_S}(s,s_0)
\hskip2mm\frac{b_{1^+_s}(s_0)}{b_{2^-_s}(s_0)}
-\frac{1}{4}\hskip2mm\alpha_0^{1S}(s,s_0)\hskip2mm I_{2^-_S 1^-_S}(s,s_0)
\hskip2mm\frac{b_{1^-_s}(s_0)}{b_{2^-_s}(s_0)}\\

\\
\hskip14mm
+\frac{3}{4}\hskip2mm\alpha_0^{0S}(s,s_0)\hskip2mm I_{2^-_S 0^+_S}(s,s_0)
\hskip2mm\frac{b_{0^+_s}(s_0)}{b_{2^-_s}(s_0)}
-\frac{1}{4}\hskip2mm\alpha_1^{1S}(s,s_0)\hskip2mm I_{2^-_S 2^-_S}(s,s_0)
\hskip2mm .\\
\end{array} \right.\eqno (85)$$

The equations of multiplet $\frac{1}{2}^-$ $(8,2)$ correspond to the
projection of orbital moment $l_z=0$. The equations of this multiplet
are similar to the case $\frac{3}{2}^-$ with replacement by the diquark
amplitudes:
$1^-\rightarrow 0^-$, $\alpha_0^1\rightarrow \alpha_0^{-0}$;
$2^-\rightarrow 1^-$, $\alpha_1^1\rightarrow \alpha_1^{-0}$;
$1^-_S\rightarrow 0^-_S$, $\alpha_0^{1S}\rightarrow \alpha_0^{-0S}$;
$2^-_S\rightarrow 1^-_S$, $\alpha_1^{1S}\rightarrow \alpha_1^{-0S}$.

\newpage
$N$ $\frac{1}{2} ^{-}$ $(8,2)$:

$$\left\{
\begin{array}{l}
\alpha_1^0(s,s_0)=\lambda-\frac{1}{4}\hskip2mm\alpha_1^0(s,s_0)
\hskip2mm I_{1^+ 1^+}(s,s_0)
+\frac{3}{4}\hskip2mm\alpha_0^{-0}(s,s_0)\hskip2mm I_{1^+ 0^-}(s,s_0)
\hskip2mm\frac{b_{0^-}(s_0)}{b_{1^+}(s_0)}\hskip5mm 1^+\\

\\
\hskip14mm
+\frac{3}{4}\hskip2mm\alpha_0^0(s,s_0)\hskip2mm I_{1^+ 0^+}(s,s_0)
\hskip2mm\frac{b_{0^+}(s_0)}{b_{1^+}(s_0)}
+\frac{3}{4}\hskip2mm\alpha_1^{-1}(s,s_0)\hskip2mm I_{1^+ 1^-}(s,s_0)
\hskip2mm\frac{b_{1^-}(s_0)}{b_{1^+}(s_0)}\\

\\
\alpha_0^{-0}(s,s_0)=\lambda+\frac{3}{4}\hskip2mm\alpha_1^0(s,s_0)
\hskip2mm I_{0^- 1^+}(s,s_0)\hskip2mm
\frac{b_{1^+}(s_0)}{b_{0^-}(s_0)}
-\frac{1}{4}\hskip2mm\alpha_0^{-0}(s,s_0)\hskip2mm I_{0^- 0^-}(s,s_0)
\hskip2.4mm 0^-\\

\\
\hskip14mm
+\frac{3}{4}\hskip2mm\alpha_0^0(s,s_0)\hskip2mm I_{0^- 0^+}(s,s_0)\hskip2mm
\frac{b_{0^+}(s_0)}{b_{0^-}(s_0)}
+\frac{3}{4}\hskip2mm\alpha_1^{-0}(s,s_0)\hskip2mm I_{0^- 1^-}(s,s_0)
\hskip2mm\frac{b_{1^-}(s_0)}{b_{0^-}(s_0)}\\

\\
\alpha_0^0(s,s_0)=\lambda+\frac{3}{4}\hskip2mm\alpha_1^0(s,s_0)
\hskip2mm I_{0^+ 1^+}(s,s_0)\hskip2mm\frac{b_{1^+}(s_0)}{b_{0^+}(s_0)}
+\frac{3}{4}\hskip2mm\alpha_0^{-0}(s,s_0)\hskip2mm I_{0^+ 0^-}(s,s_0)\cdot
\hskip2mm 0^+\\

\\
\hskip14mm
\cdot\frac{b_{0^-}(s_0)}{b_{0^+}(s_0)}-\frac{1}{4}
\hskip2mm\alpha_0^0(s,s_0)\hskip2mm I_{0^+ 0^+}(s,s_0)
+\frac{3}{4}\hskip2mm\alpha_1^{-0}(s,s_0)\hskip2mm I_{0^+ 1^-}(s,s_0)
\hskip2mm\frac{b_{1^-}(s_0)}{b_{0^+}(s_0)}\\

\\
\alpha_1^{-0}(s,s_0)=\lambda+\frac{3}{4}\hskip2mm\alpha_1^0(s,s_0)
\hskip2mm I_{1^- 1^+}(s,s_0)\hskip2mm\frac{b_{1^+}(s_0)}{b_{1^-}(s_0)}
+\frac{3}{4}\hskip2mm\alpha_0^{-0}(s,s_0)\hskip2mm I_{1^- 0^-}(s,s_0)\cdot
\hskip0mm 1^-\\

\\
\hskip14mm
\cdot\frac{b_{0^-}(s_0)}{b_{1^-}(s_0)}+\frac{3}{4}
\hskip2mm\alpha_0^0(s,s_0)\hskip2mm I_{1^- 0^+}(s,s_0)
\hskip2mm\frac{b_{0^+}(s_0)}{b_{1^-}(s_0)}
-\frac{1}{4}\hskip2mm\alpha_1^{-0}(s,s_0)\hskip2mm I_{1^- 1^-}(s,s_0)
\hskip2mm .\\
\end{array} \right.\eqno (86)$$

\newpage
$\Sigma$ $\frac{1}{2} ^{-}$ $(8,2)$:

$$\left\{
\begin{array}{l}
\alpha_1^0(s,s_0)=\lambda-\frac{1}{4}\hskip2mm\alpha_1^{0S}(s,s_0)
\hskip2mm I_{1^+ 1^+_S}(s,s_0)\hskip2mm\frac{b_{1^+_s}(s_0)}{b_{1^+}(s_0)}
\hskip40mm 1^+\\

\\
\hskip14mm
+\frac{3}{4}\hskip2mm\alpha_0^{-0S}(s,s_0)\hskip2mm I_{1^+ 0^-_S}(s,s_0)
\hskip2mm\frac{b_{0^-_s}(s_0)}{b_{1^+}(s_0)}
+\frac{3}{4}\hskip2mm\alpha_0^{0S}(s,s_0)\hskip2mm I_{1^+ 0^+_S}(s,s_0)
\hskip2mm\frac{b_{0^+_s}(s_0)}{b_{1^+}(s_0)}\\

\\
\hskip14mm
+\frac{3}{4}\hskip2mm\alpha_1^{-0S}(s,s_0)\hskip2mm I_{1^+ 1^-_S}(s,s_0)
\hskip2mm\frac{b_{1^-_s}(s_0)}{b_{1^+}(s_0)}\\

\\
\alpha_1^{0S}(s,s_0)=\lambda+\alpha_1^0(s,s_0)
\hskip2mm I_{1^+_S 1^+}(s,s_0)\hskip2mm\frac{b_{1^+}(s_0)}{b_{1^+_s}(s_0)}
\hskip45mm 1^+_s\\

\\
\hskip14mm
-\frac{5}{4}\hskip2mm\alpha_1^{0S}(s,s_0)
\hskip2mm I_{1^+_S 1^+_S}(s,s_0)
+\frac{3}{4}\hskip2mm\alpha_0^{-0S}(s,s_0)\hskip2mm I_{1^+_S 0^-_S}(s,s_0)
\hskip2mm\frac{b_{0^-_s}(s_0)}{b_{1^+_s}(s_0)}\\

\\
\hskip14mm
+\frac{3}{4}\hskip2mm\alpha_0^{0S}(s,s_0)\hskip2mm I_{1^+_S 0^+_S}(s,s_0)
\hskip2mm\frac{b_{0^+_s}(s_0)}{b_{1^+_s}(s_0)}
+\frac{3}{4}\hskip2mm\alpha_1^{-0S}(s,s_0)\hskip2mm I_{1^+_S 1^-_S}(s,s_0)
\hskip2mm\frac{b_{1^-_s}(s_0)}{b_{1^+_s}(s_0)}\\

\\
\alpha_0^{-0S}(s,s_0)=\lambda+\alpha_1^0(s,s_0)
\hskip2mm I_{0^-_S 1^+}(s,s_0)\hskip2mm\frac{b_{1^+}(s_0)}{b_{0^-_s}(s_0)}
\hskip43mm 0^-_s\\

\\
\hskip14mm
-\frac{1}{4}\hskip2mm\alpha_1^{0S}(s,s_0)
\hskip2mm I_{0^-_S 1^+_S}(s,s_0)
\hskip2mm\frac{b_{1^+_s}(s_0)}{b_{0^-_s}(s_0)}
-\frac{1}{4}\hskip2mm\alpha_0^{-0S}(s,s_0)\hskip2mm I_{0^-_S 0^-_S}(s,s_0)\\

\\
\hskip14mm
+\frac{3}{4}\hskip2mm\alpha_0^{0S}(s,s_0)\hskip2mm I_{0^-_S 0^+_S}(s,s_0)
\hskip2mm\frac{b_{0^+_s}(s_0)}{b_{0^-_s}(s_0)}
+\frac{3}{4}\hskip2mm\alpha_1^{-0S}(s,s_0)\hskip2mm I_{0^-_S 1^-_S}(s,s_0)
\hskip2mm\frac{b_{1^-_s}(s_0)}{b_{0^-_s}(s_0)}\\

\\
\alpha_0^{0S}(s,s_0)=\lambda+\alpha_1^0(s,s_0)
\hskip2mm I_{0^+_S 1^+}(s,s_0)\hskip2mm\frac{b_{1^+}(s_0)}{b_{0^+_s}(s_0)}
\hskip46mm 0^+_s\\

\\
\hskip14mm
-\frac{1}{4}\hskip2mm\alpha_1^{0S}(s,s_0)
\hskip2mm I_{0^+_S 1^+_S}(s,s_0)
\hskip2mm\frac{b_{1^+_s}(s_0)}{b_{0^+_s}(s_0)}
+\frac{3}{4}\hskip2mm\alpha_0^{-0S}(s,s_0)\hskip2mm I_{0^+_S 0^-_S}(s,s_0)
\hskip2mm\frac{b_{0^-_s}(s_0)}{b_{0^+_s}(s_0)}\\

\\
\hskip14mm
-\frac{1}{4}\hskip2mm\alpha_0^{0S}(s,s_0)\hskip2mm I_{0^+_S 0^+_S}(s,s_0)
+\frac{3}{4}\hskip2mm\alpha_1^{-0S}(s,s_0)\hskip2mm I_{0^+_S 1^-_S}(s,s_0)
\hskip2mm\frac{b_{1^-_s}(s_0)}{b_{0^+_s}(s_0)}\\

\\
\alpha_1^{-0S}(s,s_0)=\lambda+\alpha_1^0(s,s_0)
\hskip2mm I_{1^-_S 1^+}(s,s_0)\hskip2mm\frac{b_{1^+}(s_0)}{b_{1^-_s}(s_0)}
\hskip45mm 1^-_s\\

\\
\hskip14mm
-\frac{1}{4}\hskip2mm\alpha_1^{0S}(s,s_0)
\hskip2mm I_{1^-_S 1^+_S}(s,s_0)
\hskip2mm\frac{b_{1^+_s}(s_0)}{b_{1^-_s}(s_0)}
+\frac{3}{4}\hskip2mm\alpha_0^{-0S}(s,s_0)\hskip2mm I_{1^-_S 0^-_S}(s,s_0)
\hskip2mm\frac{b_{0^-_s}(s_0)}{b_{1^-_s}(s_0)}\\

\\
\hskip14mm
+\frac{3}{4}\hskip2mm\alpha_0^{0S}(s,s_0)\hskip2mm I_{1^-_S 0^+_S}(s,s_0)
\hskip2mm\frac{b_{0^+_s}(s_0)}{b_{1^-_s}(s_0)}
-\frac{1}{4}\hskip2mm\alpha_1^{-0S}(s,s_0)\hskip2mm I_{1^-_S 1^-_S}(s,s_0)
\hskip2mm .\\
\end{array} \right.\eqno (87)$$

\newpage
$\Lambda$ $\frac{1}{2} ^{-}$ $(8,2)$:

$$\left\{
\begin{array}{l}
\alpha_1^0(s,s_0)=\lambda+\frac{3}{4}\hskip2mm\alpha_1^{0S}(s,s_0)
\hskip2mm I_{1^+ 1^+_S}(s,s_0)\hskip2mm\frac{b_{1^+_s}(s_0)}{b_{1^+}(s_0)}
\hskip40mm 1^+\\

\\
\hskip14mm
-\frac{1}{4}\hskip2mm\alpha_0^{-0S}(s,s_0)\hskip2mm I_{1^+ 0^-_S}(s,s_0)
\hskip2mm\frac{b_{0^-_s}(s_0)}{b_{1^+}(s_0)}
+\frac{3}{4}\hskip2mm\alpha_0^{0S}(s,s_0)\hskip2mm I_{1^+ 0^+_S}(s,s_0)
\hskip2mm\frac{b_{0^+_s}(s_0)}{b_{1^+}(s_0)}\\

\\
\hskip14mm
+\frac{3}{4}\hskip2mm\alpha_1^{-0S}(s,s_0)\hskip2mm I_{1^+ 1^-_S}(s,s_0)
\hskip2mm\frac{b_{1^-_s}(s_0)}{b_{1^+}(s_0)}\\

\\
\alpha_1^{0S}(s,s_0)=\lambda+\alpha_1^0(s,s_0)
\hskip2mm I_{1^+_S 1^+}(s,s_0)\hskip2mm\frac{b_{1^+}(s_0)}{b_{1^+_s}(s_0)}
\hskip45mm 1^+_s\\

\\
\hskip14mm
-\frac{1}{4}\hskip2mm\alpha_1^{0S}(s,s_0)
\hskip2mm I_{1^+_S 1^+_S}(s,s_0)
-\frac{1}{4}\hskip2mm\alpha_0^{-0S}(s,s_0)\hskip2mm I_{1^+_S 0^-_S}(s,s_0)
\hskip2mm\frac{b_{0^-_s}(s_0)}{b_{1^+_s}(s_0)}\\

\\
\hskip14mm
+\frac{3}{4}\hskip2mm\alpha_0^{0S}(s,s_0)\hskip2mm I_{1^+_S 0^+_S}(s,s_0)
\hskip2mm\frac{b_{0^+_s}(s_0)}{b_{1^+_s}(s_0)}
+\frac{3}{4}\hskip2mm\alpha_1^{-0S}(s,s_0)\hskip2mm I_{1^+_S 1^-_S}(s,s_0)
\hskip2mm\frac{b_{1^-_s}(s_0)}{b_{1^+_s}(s_0)}\\

\\
\alpha_0^{-0S}(s,s_0)=\lambda+\alpha_1^0(s,s_0)
\hskip2mm I_{0^-_S 1^+}(s,s_0)\hskip2mm\frac{b_{1^+}(s_0)}{b_{0^-_s}(s_0)}
\hskip43mm 0^-_s\\

\\
\hskip14mm
+\frac{3}{4}\hskip2mm\alpha_1^{0S}(s,s_0)
\hskip2mm I_{0^-_S 1^+_S}(s,s_0)
\hskip2mm\frac{b_{1^+_s}(s_0)}{b_{0^-_s}(s_0)}
-\frac{5}{4}\hskip2mm\alpha_0^{-0S}(s,s_0)\hskip2mm I_{0^-_S 0^-_S}(s,s_0)\\

\\
\hskip14mm
+\frac{3}{4}\hskip2mm\alpha_0^{0S}(s,s_0)\hskip2mm I_{0^-_S 0^+_S}(s,s_0)
\hskip2mm\frac{b_{0^+_s}(s_0)}{b_{0^-_s}(s_0)}
+\frac{3}{4}\hskip2mm\alpha_1^{-0S}(s,s_0)\hskip2mm I_{0^-_S 1^-_S}(s,s_0)
\hskip2mm\frac{b_{1^-_s}(s_0)}{b_{0^-_s}(s_0)}\\

\\
\alpha_0^{0S}(s,s_0)=\lambda+\alpha_1^0(s,s_0)
\hskip2mm I_{0^+_S 1^+}(s,s_0)\hskip2mm\frac{b_{1^+}(s_0)}{b_{0^+_s}(s_0)}
\hskip46mm 0^+_s\\

\\
\hskip14mm
+\frac{3}{4}\hskip2mm\alpha_1^{0S}(s,s_0)
\hskip2mm I_{0^+_S 1^+_S}(s,s_0)
\hskip2mm\frac{b_{1^+_s}(s_0)}{b_{0^+_s}(s_0)}
-\frac{1}{4}\hskip2mm\alpha_0^{-0S}(s,s_0)\hskip2mm I_{0^+_S 0^-_S}(s,s_0)
\hskip2mm\frac{b_{0^-_s}(s_0)}{b_{0^+_s}(s_0)}\\

\\
\hskip14mm
-\frac{1}{4}\hskip2mm\alpha_0^{0S}(s,s_0)\hskip2mm I_{0^+_S 0^+_S}(s,s_0)
+\frac{3}{4}\hskip2mm\alpha_1^{-0S}(s,s_0)\hskip2mm I_{0^+_S 1^-_S}(s,s_0)
\hskip2mm\frac{b_{1^-_s}(s_0)}{b_{0^+_s}(s_0)}\\

\\
\alpha_1^{-0S}(s,s_0)=\lambda+\alpha_1^0(s,s_0)
\hskip2mm I_{1^-_S 1^+}(s,s_0)\hskip2mm\frac{b_{1^+}(s_0)}{b_{1^-_s}(s_0)}
\hskip45mm 1^-_s\\

\\
\hskip14mm
+\frac{3}{4}\hskip2mm\alpha_1^{0S}(s,s_0)
\hskip2mm I_{1^-_S 1^+_S}(s,s_0)
\hskip2mm\frac{b_{1^+_s}(s_0)}{b_{1^-_s}(s_0)}
-\frac{1}{4}\hskip2mm\alpha_0^{-0S}(s,s_0)\hskip2mm I_{1^-_S 0^-_S}(s,s_0)
\hskip2mm\frac{b_{0^-_s}(s_0)}{b_{1^-_s}(s_0)}\\

\\
\hskip14mm
+\frac{3}{4}\hskip2mm\alpha_0^{0S}(s,s_0)\hskip2mm I_{1^-_S 0^+_S}(s,s_0)
\hskip2mm\frac{b_{0^+_s}(s_0)}{b_{1^-_s}(s_0)}
-\frac{1}{4}\hskip2mm\alpha_1^{-0S}(s,s_0)\hskip2mm I_{1^-_S 1^-_S}(s,s_0)
\hskip2mm .\\
\end{array} \right.\eqno (88)$$

\newpage
{\bf C3. The equations of ${\bf (8,4)}$ multiplet.}
\vskip 5mm

At first we consider the multiplet $\frac{5}{2} ^{-}$ $(8,4)$. In this
case the spin $S=\frac{3}{2}$ and $l_z=+1$.

$N$ $\frac{5}{2} ^{-}$ $(8,4)$:

$$\left\{
\begin{array}{l}
\alpha_1^0(s,s_0)=\lambda+\frac{1}{2}\hskip2mm\alpha_1^0(s,s_0)
\hskip2mm I_{1^+ 1^+}(s,s_0)\hskip40mm 1^+\\
\hskip25mm
+\frac{3}{2}\hskip2mm\alpha_1^1(s,s_0)\hskip2mm I_{1^+ 2^-}(s,s_0)\hskip2mm
\frac{b_{2^-}(s_0)}{b_{1^+}(s_0)}\\
\alpha_1^1(s,s_0)=\lambda+\frac{3}{2}\hskip2mm\alpha_1^0(s,s_0)
\hskip2mm I_{2^- 1^+}(s,s_0)
\hskip2mm\frac{b_{1^+}(s_0)}{b_{2^-}(s_0)}\hskip27mm 2^-\\
\hskip25mm
+\frac{1}{2}\hskip2mm\alpha_1^1(s,s_0)\hskip2mm I_{2^- 2^-}(s,s_0)
\hskip2mm .\\
\end{array} \right.\eqno (89)$$

$\Sigma$ $\frac{5}{2} ^{-}$ $(8,4)$:

$$\left\{
\begin{array}{l}
\alpha_1^0(s,s_0)=\lambda+\frac{1}{2}\hskip2mm\alpha_1^{0S}(s,s_0)
\hskip2mm I_{1^+ 1^+_S}(s,s_0)
\hskip2mm\frac{b_{1^+_s}(s_0)}{b_{1^+}(s_0)}\hskip40mm 1^+\\
\hskip25mm
+\frac{3}{2}\hskip2mm\alpha_1^{1S}(s,s_0)\hskip2mm I_{1^+ 2^-_S}(s,s_0)
\hskip2mm\frac{b_{2^-_s}(s_0)}{b_{1^+}(s_0)}\\
\alpha_1^{0S}(s,s_0)=\lambda
+\alpha_1^0(s,s_0)\hskip2mm I_{1^+_S 1^+}(s,s_0)
\hskip2mm\frac{b_{1^+}(s_0)}{b_{1^+_s}(s_0)}\hskip45mm 1^+_S\\
\hskip25mm
-\frac{1}{2}\hskip2mm\alpha_1^{0S}(s,s_0)
\hskip2mm I_{1^+_S 1^+_S}(s,s_0)
+\frac{3}{2}\hskip2mm\alpha_1^{1S}(s,s_0)\hskip2mm I_{1^+_S 2^-_S}(s,s_0)
\hskip2mm\frac{b_{2^-_s}(s_0)}{b_{1^+_s}(s_0)}\\
\alpha_1^{1S}(s,s_0)=\lambda+\alpha_1^0(s,s_0)\hskip2mm I_{2^-_S 1^+}(s,s_0)
\hskip2mm\frac{b_{1^+}(s_0)}{b_{2^-_s}(s_0)}\hskip45mm 2^-_S\\
\hskip25mm
+\frac{1}{2}\hskip2mm\alpha_1^{0S}(s,s_0)
\hskip2mm I_{2^-_S 1^+_S}(s,s_0)
\hskip2mm\frac{b_{1^+_s}(s_0)}{b_{2^-_s}(s_0)}
+\frac{1}{2}\hskip2mm\alpha_1^{1S}(s,s_0)\hskip2mm I_{2^-_S 2^-_S}(s,s_0)
\hskip2mm .\\
\end{array} \right.\eqno (90)$$

$\Lambda$ $\frac{5}{2} ^{-}$ $(8,4)$:

$$\left\{
\begin{array}{l}
\alpha_1^1(s,s_0)=\lambda+\frac{3}{2}\hskip2mm\alpha_1^{0S}(s,s_0)
\hskip2mm I_{2^- 1^+_S}(s,s_0)
\hskip2mm\frac{b_{1^+_s}(s_0)}{b_{2^-}(s_0)}\hskip40mm 2^-\\
\hskip25mm
+\frac{1}{2}\hskip2mm\alpha_1^{1S}(s,s_0)\hskip2mm I_{2^- 2^-_S}(s,s_0)
\hskip2mm\frac{b_{2^-_s}(s_0)}{b_{2^-}(s_0)}\\
\alpha_1^{0S}(s,s_0)=\lambda
+\alpha_1^1(s,s_0)\hskip2mm I_{1^+_S 2^-}(s,s_0)
\hskip2mm\frac{b_{2^-}(s_0)}{b_{1^+_s}(s_0)}\hskip45mm 1^+_S\\
\hskip25mm
+\frac{2}{3}\hskip2mm\alpha_1^{0S}(s,s_0)
\hskip2mm I_{1^+_S 1^+_S}(s,s_0)
+\frac{1}{3}\hskip2mm\alpha_1^{1S}(s,s_0)\hskip2mm I_{1^+_S 2^-_S}(s,s_0)
\hskip2mm\frac{b_{2^-_s}(s_0)}{b_{1^+_s}(s_0)}\\
\alpha_1^{1S}(s,s_0)=\lambda+\alpha_1^1(s,s_0)\hskip2mm I_{2^-_S 2^-}(s,s_0)
\hskip2mm\frac{b_{2^-}(s_0)}{b_{2^-_s}(s_0)}\hskip45mm 2^-_S\\
\hskip25mm
+\hskip2mm\alpha_1^{0S}(s,s_0)\hskip2mm I_{2^-_S 1^+_S}(s,s_0)
\hskip2mm\frac{b_{1^+_s}(s_0)}{b_{2^-_s}(s_0)}
\hskip2mm .\\
\end{array} \right.\eqno (91)$$

We derive the equations for the multiplet $\frac{3}{2} ^{-}$ $(8,4)$.
One use the spin $S=\frac{3}{2}$ and $l_z=0$.

$N$ $\frac{3}{2} ^{-}$ $(8,4)$:

$$\left\{
\begin{array}{l}
\alpha_1^0(s,s_0)=\lambda+\frac{1}{2}\hskip2mm\alpha_1^0(s,s_0)
\hskip2mm I_{1^+ 1^+}(s,s_0)\hskip43mm 1^+\\
\hskip25mm
+\frac{3}{2}\hskip2mm\alpha_1^{-0}(s,s_0)\hskip2mm I_{1^+ 1^-}(s,s_0)
\hskip2mm\frac{b_{1^-}(s_0)}{b_{1^+}(s_0)}\\
\alpha_1^{-0}(s,s_0)=\lambda+\frac{3}{2}\hskip2mm\alpha_1^0(s,s_0)
\hskip2mm I_{1^- 1^+}(s,s_0)
\hskip2mm\frac{b_{1^+}(s_0)}{b_{1^-}(s_0)}\hskip28mm 1^-\\
\hskip25mm
+\frac{1}{2}\hskip2mm\alpha_1^{-0}(s,s_0)\hskip2mm I_{1^- 1^-}(s,s_0)
\hskip2mm .\\
\end{array} \right.\eqno (92)$$

\newpage
$\Sigma$ $\frac{3}{2} ^{-}$ $(8,4)$:

$$\left\{
\begin{array}{l}
\alpha_1^0(s,s_0)=\lambda+\frac{1}{2}\hskip2mm\alpha_1^{0S}(s,s_0)
\hskip2mm I_{1^+ 1^+_S}(s,s_0)
\hskip2mm\frac{b_{1^+_s}(s_0)}{b_{1^+}(s_0)}\hskip40mm 1^+\\
\hskip25mm
+\frac{3}{2}\hskip2mm\alpha_1^{-0S}(s,s_0)\hskip2mm I_{1^+ 1^-_S}(s,s_0)
\hskip2mm\frac{b_{1^-_s}(s_0)}{b_{1^+}(s_0)}\\
\alpha_1^{0S}(s,s_0)=\lambda
+\alpha_1^0(s,s_0)\hskip2mm I_{1^+_S 1^+}(s,s_0)
\hskip2mm\frac{b_{1^+}(s_0)}{b_{1^+_s}(s_0)}\hskip44mm 1^+_S\\
\hskip25mm
-\frac{1}{2}\hskip2mm\alpha_1^{0S}(s,s_0)
\hskip2mm I_{1^+_S 1^+_S}(s,s_0)
+\frac{3}{2}\hskip2mm\alpha_1^{-0S}(s,s_0)\hskip2mm I_{1^+_S 1^-_S}(s,s_0)
\hskip2mm\frac{b_{1^-_s}(s_0)}{b_{1^+_s}(s_0)}\\
\alpha_1^{-0S}(s,s_0)=\lambda+\alpha_1^0(s,s_0)
\hskip2mm I_{1^-_S 1^+}(s,s_0)
\hskip2mm\frac{b_{1^+}(s_0)}{b_{1^-_s}(s_0)}\hskip42mm 1^-_S\\
\hskip25mm
+\frac{1}{2}\hskip2mm\alpha_1^{0S}(s,s_0)
\hskip2mm I_{1^-_S 1^+_S}(s,s_0)
\hskip2mm\frac{b_{1^+_s}(s_0)}{b_{1^-_s}(s_0)}
+\frac{1}{2}\hskip2mm\alpha_1^{-0S}(s,s_0)\hskip2mm I_{1^-_S 1^-_S}(s,s_0)
\hskip2mm .\\
\end{array} \right.\eqno (93)$$

$\Lambda$ $\frac{3}{2} ^{-}$ $(8,4)$:

$$\left\{
\begin{array}{l}
\alpha_1^{-0}(s,s_0)=\lambda+\frac{3}{2}\hskip2mm\alpha_1^{0S}(s,s_0)
\hskip2mm I_{1^- 1^+_S}(s,s_0)
\hskip2mm\frac{b_{1^+_s}(s_0)}{b_{1^-}(s_0)}\hskip40mm 1^-\\
\hskip25mm
+\frac{1}{2}\hskip2mm\alpha_1^{-0S}(s,s_0)\hskip2mm I_{1^- 1^-_S}(s,s_0)
\hskip2mm\frac{b_{1^-_s}(s_0)}{b_{1^-}(s_0)}\\
\alpha_1^{0S}(s,s_0)=\lambda
+\alpha_1^{-0}(s,s_0)\hskip2mm I_{1^+_S 1^-}(s,s_0)
\hskip2mm\frac{b_{1^-}(s_0)}{b_{1^+_s}(s_0)}\hskip45mm 1^+_S\\
\hskip25mm
+\frac{2}{3}\hskip2mm\alpha_1^{0S}(s,s_0)
\hskip2mm I_{1^+_S 1^+_S}(s,s_0)
+\frac{1}{3}\hskip2mm\alpha_1^{-0S}(s,s_0)\hskip2mm I_{1^+_S 1^-_S}(s,s_0)
\hskip2mm\frac{b_{1^-_s}(s_0)}{b_{1^+_s}(s_0)}\\
\alpha_1^{-0S}(s,s_0)=\lambda+\alpha_1^{-0}(s,s_0)
\hskip2mm I_{1^-_S 1^-}(s,s_0)
\hskip2mm\frac{b_{1^-}(s_0)}{b_{1^-_s}(s_0)}\hskip42mm 1^-_S\\
\hskip25mm
+\hskip2mm\alpha_1^{0S}(s,s_0)\hskip2mm I_{1^-_S 1^+_S}(s,s_0)
\hskip2mm\frac{b_{1^+_s}(s_0)}{b_{1^-_s}(s_0)}
\hskip2mm .\\
\end{array} \right.\eqno (94)$$

We use the spin $S=\frac{3}{2}$ and the moment projection $l_z=-1$ and
obtain the equations of multiplet $\frac{1}{2} ^{-}$ $(8,4)$.

$N$ $\frac{1}{2} ^{-}$ $(8,4)$:

$$\left\{
\begin{array}{l}
\alpha_1^0(s,s_0)=\lambda+\frac{1}{2}\hskip2mm\alpha_1^0(s,s_0)
\hskip2mm I_{1^+ 1^+}(s,s_0)\hskip43mm 1^+\\
\hskip25mm
+\frac{3}{2}\hskip2mm\alpha_1^{-1}(s,s_0)\hskip2mm I_{1^+ 0^-}(s,s_0)
\hskip2mm\frac{b_{0^-}(s_0)}{b_{1^+}(s_0)}\\
\alpha_1^{-1}(s,s_0)=\lambda+\frac{3}{2}\hskip2mm\alpha_1^0(s,s_0)
\hskip2mm I_{0^- 1^+}(s,s_0)
\hskip2mm\frac{b_{1^+}(s_0)}{b_{0^-}(s_0)}\hskip28mm 0^-\\
\hskip25mm
+\frac{1}{2}\hskip2mm\alpha_1^{-1}(s,s_0)\hskip2mm I_{0^- 0^-}(s,s_0)
\hskip2mm .\\
\end{array} \right.\eqno (95)$$

$\Sigma$ $\frac{1}{2} ^{-}$ $(8,4)$:

$$\left\{
\begin{array}{l}
\alpha_1^0(s,s_0)=\lambda+\frac{1}{2}\hskip2mm\alpha_1^{0S}(s,s_0)
\hskip2mm I_{1^+ 1^+_S}(s,s_0)
\hskip2mm\frac{b_{1^+_s}(s_0)}{b_{1^+}(s_0)}\hskip40mm 1^+\\
\hskip25mm
+\frac{3}{2}\hskip2mm\alpha_1^{-1S}(s,s_0)\hskip2mm I_{1^+ 0^-_S}(s,s_0)
\hskip2mm\frac{b_{0^-_s}(s_0)}{b_{1^+}(s_0)}\\
\alpha_1^{0S}(s,s_0)=\lambda
+\alpha_1^0(s,s_0)\hskip2mm I_{1^+_S 1^+}(s,s_0)
\hskip2mm\frac{b_{1^+}(s_0)}{b_{1^+_s}(s_0)}\hskip45mm 1^+_S\\
\hskip25mm
-\frac{1}{2}\hskip2mm\alpha_1^{0S}(s,s_0)
\hskip2mm I_{1^+_S 1^+_S}(s,s_0)
+\frac{3}{2}\hskip2mm\alpha_1^{-1S}(s,s_0)\hskip2mm I_{1^+_S 0^-_S}(s,s_0)
\hskip2mm\frac{b_{0^-_s}(s_0)}{b_{1^+_s}(s_0)}\\
\alpha_1^{-1S}(s,s_0)=\lambda+\alpha_1^0(s,s_0)
\hskip2mm I_{0^-_S 1^+}(s,s_0)
\hskip2mm\frac{b_{1^+}(s_0)}{b_{0^-_s}(s_0)}\hskip43mm 0^-_S\\
\hskip25mm
+\frac{1}{2}\hskip2mm\alpha_1^{0S}(s,s_0)
\hskip2mm I_{0^-_S 1^+_S}(s,s_0)
\hskip2mm\frac{b_{1^+_s}(s_0)}{b_{0^-_s}(s_0)}
+\frac{1}{2}\hskip2mm\alpha_1^{-1S}(s,s_0)\hskip2mm I_{0^-_S 0^-_S}(s,s_0)
\hskip2mm .\\
\end{array} \right.\eqno (96)$$

$\Lambda$ $\frac{1}{2} ^{-}$ $(8,4)$:

$$\left\{
\begin{array}{l}
\alpha_1^{-1}(s,s_0)=\lambda+\frac{3}{2}\hskip2mm\alpha_1^{0S}(s,s_0)
\hskip2mm I_{0^- 1^+_S}(s,s_0)
\hskip2mm\frac{b_{1^+_s}(s_0)}{b_{0^-}(s_0)}\hskip40mm 0^-\\
\hskip25mm
+\frac{1}{2}\hskip2mm\alpha_1^{-1S}(s,s_0)\hskip2mm I_{0^- 0^-_S}(s,s_0)
\hskip2mm\frac{b_{0^-_s}(s_0)}{b_{0^-}(s_0)}\\
\alpha_1^{0S}(s,s_0)=\lambda
+\alpha_1^{-1}(s,s_0)\hskip2mm I_{1^+_S 0^-}(s,s_0)
\hskip2mm\frac{b_{0^-}(s_0)}{b_{1^+_s}(s_0)}\hskip45mm 1^+_S\\
\hskip25mm
+\frac{2}{3}\hskip2mm\alpha_1^{0S}(s,s_0)
\hskip2mm I_{1^+_S 1^+_S}(s,s_0)
+\frac{1}{3}\hskip2mm\alpha_1^{-1S}(s,s_0)\hskip2mm I_{1^+_S 0^-_S}(s,s_0)
\hskip2mm\frac{b_{0^-_s}(s_0)}{b_{1^+_s}(s_0)}\\
\alpha_1^{-1S}(s,s_0)=\lambda+\alpha_1^{-1}(s,s_0)
\hskip2mm I_{0^-_S 0^-}(s,s_0)
\hskip2mm\frac{b_{0^-}(s_0)}{b_{0^-_s}(s_0)}\hskip43mm 0^-_S\\
\hskip25mm
+\hskip2mm\alpha_1^{0S}(s,s_0)\hskip2mm I_{0^-_S 1^+_S}(s,s_0)
\hskip2mm\frac{b_{1^+_s}(s_0)}{b_{0^-_s}(s_0)}
\hskip2mm .\\
\end{array} \right.\eqno (97)$$

\vskip 5mm
{\bf C4. The equations of ${\bf (1,2)}$ singlet.}
\vskip 5mm

$\Lambda$ $\frac{3}{2} ^{-}$ $(1,2)$:

$$\left\{
\begin{array}{l}
\alpha_0^0(s,s_0)=\lambda+\frac{1}{2}\hskip2mm\alpha_0^{0S}(s,s_0)
\hskip2mm I_{0^+ 0^+_S}(s,s_0)
\hskip2mm\frac{b_{0^+_s}(s_0)}{b_{0^+}(s_0)}\hskip40mm 0^+\\
\hskip25mm
+\frac{3}{2}\hskip2mm\alpha_1^{1S}(s,s_0)\hskip2mm I_{0^+ 2^-_S}(s,s_0)
\hskip2mm\frac{b_{2^-_s}(s_0)}{b_{0^+}(s_0)}\\
\alpha_0^{0S}(s,s_0)=\lambda
+\alpha_0^0(s,s_0)\hskip2mm I_{0^+_S 0^+}(s,s_0)
\hskip2mm\frac{b_{0^+}(s_0)}{b_{0^+_s}(s_0)}\hskip45mm 0^+_S\\
\hskip25mm
-\frac{1}{2}\hskip2mm\alpha_0^{0S}(s,s_0)
\hskip2mm I_{0^+_S 0^+_S}(s,s_0)
+\frac{3}{2}\hskip2mm\alpha_1^{1S}(s,s_0)\hskip2mm I_{0^+_S 2^-_S}(s,s_0)
\hskip2mm\frac{b_{2^-_s}(s_0)}{b_{0^+_s}(s_0)}\\
\alpha_1^{1S}(s,s_0)=\lambda+\alpha_0^0(s,s_0)\hskip2mm I_{2^-_S 0^+}(s,s_0)
\hskip2mm\frac{b_{0^+}(s_0)}{b_{2^-_s}(s_0)}\hskip45mm 2^-_S\\
\hskip25mm
+\frac{1}{2}\hskip2mm\alpha_0^{0S}(s,s_0)
\hskip2mm I_{2^-_S 0^+_S}(s,s_0)
\hskip2mm\frac{b_{0^+_s}(s_0)}{b_{2^-_s}(s_0)}
+\frac{1}{2}\hskip2mm\alpha_1^{1S}(s,s_0)\hskip2mm I_{2^-_S 2^-_S}(s,s_0)
\hskip2mm .\\
\end{array} \right.\eqno (98)$$

$\Lambda$ $\frac{1}{2} ^{-}$ $(1,2)$:

$$\left\{
\begin{array}{l}
\alpha_0^0(s,s_0)=\lambda+\frac{1}{2}\hskip2mm\alpha_0^{0S}(s,s_0)
\hskip2mm I_{0^+ 0^+_S}(s,s_0)
\hskip2mm\frac{b_{0^+_s}(s_0)}{b_{0^+}(s_0)}\hskip40mm 0^+\\
\hskip25mm
+\frac{3}{2}\hskip2mm\alpha_1^{-0S}(s,s_0)\hskip2mm I_{0^+ 1^-_S}(s,s_0)
\hskip2mm\frac{b_{1^-_s}(s_0)}{b_{0^+}(s_0)}\\
\alpha_0^{0S}(s,s_0)=\lambda
+\alpha_0^0(s,s_0)\hskip2mm I_{0^+_S 0^+}(s,s_0)
\hskip2mm\frac{b_{0^+}(s_0)}{b_{0^+_s}(s_0)}\hskip45mm 0^+_S\\
\hskip25mm
-\frac{1}{2}\hskip2mm\alpha_0^{0S}(s,s_0)
\hskip2mm I_{0^+_S 0^+_S}(s,s_0)
+\frac{3}{2}\hskip2mm\alpha_1^{-0S}(s,s_0)\hskip2mm I_{0^+_S 1^-_S}(s,s_0)
\hskip2mm\frac{b_{1^-_s}(s_0)}{b_{0^+_s}(s_0)}\\
\alpha_1^{-0S}(s,s_0)=\lambda+\alpha_0^0(s,s_0)
\hskip2mm I_{1^-_S 0^+}(s,s_0)
\hskip2mm\frac{b_{0^+}(s_0)}{b_{1^-_s}(s_0)}\hskip42mm 1^-_S\\
\hskip25mm
+\frac{1}{2}\hskip2mm\alpha_0^{0S}(s,s_0)
\hskip2mm I_{1^-_S 0^+_S}(s,s_0)
\hskip2mm\frac{b_{0^+_s}(s_0)}{b_{1^-_s}(s_0)}
+\frac{1}{2}\hskip2mm\alpha_1^{-0S}(s,s_0)\hskip2mm I_{1^-_S 1^-_S}(s,s_0)
\hskip2mm .\\
\end{array} \right.\eqno (99)$$

\newpage
{\LARGE
{\bf
References.}}
\vskip5ex

\noindent
1. G.'t Hooft, Nucl. Phys. B{\bf 72} (1974) 461.

\noindent
2. E. Witten, Nucl. Phys. B{\bf 160} (1979) 57.

\noindent
3. J.L. Gervais and B. Sakita, Phys. Rev. Lett. {\bf 52} (1984) 87.

\noindent
4. R. Dashen and A.V. Manohar, Phys. Lett. B{\bf 315} (1993) 425.

\noindent
5. R. Dashen, E. Jenkins and A.V. Manohar, Phys. Rev. D{\bf 49} (1994) 4713.

\noindent
6. R. Dashen, E. Jenkins and A.V. Manohar, Phys. Rev. D{\bf 51} (1995) 3697.

\noindent
7. C.D. Carone, H. Georgi and S. Osofsky, Phys. Lett. B{\bf 322} (1994) 227.

\noindent
8. M.A. Luty and J. March-Russell, Nucl. Phys. B{\bf 426} (1994) 71

\noindent
9. E. Jenkins, Phys. Lett. B{\bf 315} (1993) 441.

\noindent
10. E. Jenkins and R.F. Lebed, Phys. Rev. D{\bf 52} (1995) 282.

\noindent
11. J. Dai, R. Dashen, E. Jenkins, and A. V. Manohar,
Phys. Rev. D{\bf 53} (1996) 273.

\noindent
12. N. Matagne and Fl. Stancu, Phys. Rev. 2005 D{\bf 71} 015710.

\noindent
13. N. Matagne and Fl. Stancu hep-ph/0610099.

\noindent
14. I.J.R. Aitchison, J. Phys. G{\bf 3} (1977) 121.

\noindent
15. J.J. Brehm, Ann. Phys. (N.Y.) {\bf 108} (1977) 454.

\noindent
16. I.J.R. Aitchison and J.J. Brehm, Phys. Rev. D{\bf 17} (1978) 3072.

\noindent
17. I.J.R. Aitchison and J.J. Brehm, Phys. Rev. D{\bf 20} (1979) 1119, 1131.

\noindent
18. J.J. Brehm, Phys. Rev. D{\bf 21} (1980) 718.

\noindent
19. S.M. Gerasyuta, Yad. Fiz. {\bf 55} (1992) 3030.

\noindent
20. S.M. Gerasyuta, Z. Phys. C{\bf 60} (1993) 683.

\noindent
21. F.E. Close, {\it An introduction to quarks and partons}, Academic
Press London

New York San Francisco, (1979) P. 438.

\noindent
22. A.De Rujula, H.Georgi and S.L.Glashow, Phys. Rev. D{\bf 12} (1975) 147.

\noindent
23. V.V. Anisovich, S.M. Gerasyuta and A.V. Sarantsev,
Int. J. Mod. Phis. A{\bf 6} (1991)

625.

\noindent
24. G. Chew and S. Mandelstam, Phys. Rev. {\bf 119} (1960) 467.

\noindent
25. V.V. Anisovich and A.A. Anselm, Usp. Fiz. Nauk {\bf 88} (1966) 287.

\noindent
26. V.V. Anisovich and S.M. Gerasyuta, Yad. Fiz. {\bf 44} (1986) 174.

\noindent
27. S.M. Gerasyuta and I.V. Keltuyala, Yad. Fiz. {\bf 54} (1991) 793.

\noindent
28. W.M. Yao et al. (Particle Data Group), J. Phys. G{\bf 33} (2006) 1.

\noindent
29. S.M. Gerasyuta and I.V. Kochkin, Int. J. Mod. Phys. E{\bf 15} (2006) 71.

\noindent
30. S.M. Gerasyuta and I.V. Kochkin, Phys. Rev. D{\bf 66} (2002) 116001.

\noindent
31. A.J. Hey and R.L. Kelly, Phys. Rept. {\bf 96} (1983) 71.

\noindent
32. S. Capstick and W. Roberts, nucl-th/000828.

\noindent
33. S. Capstick and N. Isgur, Phys. Rev. D{\bf 34} (1986) 2809.

\noindent
34. S.M. Gerasyuta and D.V. Ivanov, Vest. SPb University
Ser. 4 {\bf 11} (1996) 3.

\newpage
\vskip60pt
\begin{picture}(600,60)
\put(-10,40){\line(1,0){33}}
\put(-10,50){\line(1,0){28}}
\put(-10,60){\line(1,0){33}}
\put(19,46){\line(1,1){15}}
\put(22,41){\line(1,1){17}}
\put(27.5,38.5){\line(1,1){14}}
\put(41,56){\vector(2,1){28}}
\put(42.5,50){\vector(1,0){35}}
\put(41,44){\vector(2,-1){28}}
\put(30,50){\circle{25}}
\put(70,78){$1$}
\put(70,55){$2$}
\put(70,20){$3$}
\put(87,47){$=$}
\put(107,53){\line(1,0){28}}
\put(107,50){\line(1,0){28}}
\put(107,47){\line(1,0){28}}
\put(135,53){\vector(2,1){28}}
\put(135,50){\vector(1,0){35}}
\put(135,47){\vector(2,-1){28}}
\put(163,78){$1$}
\put(163,55){$2$}
\put(163,20){$3$}
\put(180,47){$+$}
\put(200,40){\line(1,0){33}}
\put(200,50){\line(1,0){28}}
\put(200,60){\line(1,0){33}}
\put(229,46){\line(1,1){15}}
\put(232,41){\line(1,1){17}}
\put(237.5,38.5){\line(1,1){14}}
\put(251,44){\vector(2,-1){28}}
\put(240,50){\circle{25}}
\put(268,54){\oval(33,33)[tl]}
\put(252,70){\oval(33,33)[br]}
\put(269,71){\vector(2,3){15}}
\put(269,71){\vector(2,-1){24}}
\put(287,95){$1$}
\put(295,65){$2$}
\put(280,20){$3$}
\put(300,47){$+$}
\end{picture}

\vskip60pt
\begin{picture}(600,60)
\put(90,47){$+$}
\put(110,40){\line(1,0){33}}
\put(110,50){\line(1,0){28}}
\put(110,60){\line(1,0){33}}
\put(139,46){\line(1,1){15}}
\put(142,41){\line(1,1){17}}
\put(147.5,38.5){\line(1,1){14}}
\put(161,44){\vector(2,-1){28}}
\put(150,50){\circle{25}}
\put(178,54){\oval(33,33)[tl]}
\put(162,70){\oval(33,33)[br]}
\put(179,71){\vector(2,3){15}}
\put(179,71){\vector(2,-1){24}}
\put(197,95){$1$}
\put(205,65){$3$}
\put(190,20){$2$}
\put(210,47){$+$}
\put(230,40){\line(1,0){33}}
\put(230,50){\line(1,0){28}}
\put(230,60){\line(1,0){33}}
\put(259,46){\line(1,1){15}}
\put(262,41){\line(1,1){17}}
\put(267.5,38.5){\line(1,1){14}}
\put(281,44){\vector(2,-1){28}}
\put(270,50){\circle{25}}
\put(298,54){\oval(33,33)[tl]}
\put(282,70){\oval(33,33)[br]}
\put(299,71){\vector(2,3){15}}
\put(299,71){\vector(2,-1){24}}
\put(317,95){$2$}
\put(325,65){$3$}
\put(310,20){$1$}
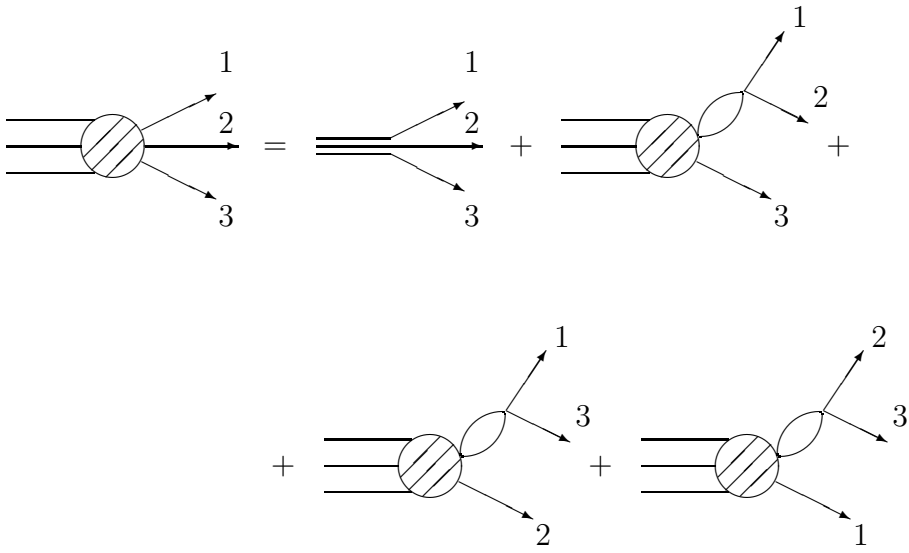
\put(-10,0){{\large Fig.1. The contribution of diagrams at the last pair
of the interacting particles.}}
\end{picture}

\vskip60pt

\vskip60pt
\begin{picture}(600,60)
\put(-10,40){\line(1,0){33}}
\put(-10,50){\line(1,0){28}}
\put(-10,60){\line(1,0){33}}
\put(19,46){\line(1,1){15}}
\put(22,41){\line(1,1){17}}
\put(27.5,38.5){\line(1,1){14}}
\put(41,44){\vector(2,-1){28}}
\put(30,50){\circle{25}}
\put(58,54){\oval(33,33)[tl]}
\put(42,70){\oval(33,33)[br]}
\put(59,71){\vector(2,3){15}}
\put(59,71){\vector(2,-1){24}}
\put(77,95){$1$}
\put(85,65){$2$}
\put(70,20){$3$}
\put(90,47){$=$}
\put(110,52){\line(1,0){28}}
\put(110,50){\line(1,0){28}}
\put(110,48){\line(1,0){28}}
\put(154,51){\oval(33,33)[tl]}
\put(138,67){\oval(33,33)[br]}
\put(155,67){\vector(2,3){15}}
\put(155,67){\vector(2,-1){24}}
\put(139,49){\vector(2,-1){28}}
\put(173,91){$1$}
\put(181,61){$2$}
\put(168,25){$3$}
\put(190,47){$+$}
\put(210,40){\line(1,0){33}}
\put(210,50){\line(1,0){28}}
\put(210,60){\line(1,0){33}}
\put(239,46){\line(1,1){15}}
\put(242,41){\line(1,1){17}}
\put(247.5,38.5){\line(1,1){14}}
\put(261,44){\vector(1,0){43}}
\put(250,50){\circle{25}}
\put(278,54){\oval(33,33)[tl]}
\put(262,70){\oval(33,33)[br]}
\put(279,71){\vector(2,3){15}}
\put(279,71){\vector(1,-1){25}}
\put(297,95){$3$}
\put(300,60){$1$}
\put(290,27){$2$}
\put(305,29){\oval(33,33)[tr]}
\put(321,45){\oval(33,33)[bl]}
\put(323,29){\vector(2,3){15}}
\put(323,29){\vector(2,-1){24}}
\put(341,53){$1$}
\put(333,7){$2$}
\put(360,47){$+$}
\end{picture}

\vskip60pt
\begin{picture}(600,60)
\put(190,47){$+$}
\put(210,40){\line(1,0){33}}
\put(210,50){\line(1,0){28}}
\put(210,60){\line(1,0){33}}
\put(239,46){\line(1,1){15}}
\put(242,41){\line(1,1){17}}
\put(247.5,38.5){\line(1,1){14}}
\put(261,44){\vector(1,0){43}}
\put(250,50){\circle{25}}
\put(278,54){\oval(33,33)[tl]}
\put(262,70){\oval(33,33)[br]}
\put(279,71){\vector(2,3){15}}
\put(279,71){\vector(1,-1){25}}
\put(297,95){$3$}
\put(300,60){$2$}
\put(290,27){$1$}
\put(305,29){\oval(33,33)[tr]}
\put(321,45){\oval(33,33)[bl]}
\put(323,29){\vector(2,3){15}}
\put(323,29){\vector(2,-1){24}}
\put(341,53){$2$}
\put(333,7){$1$}
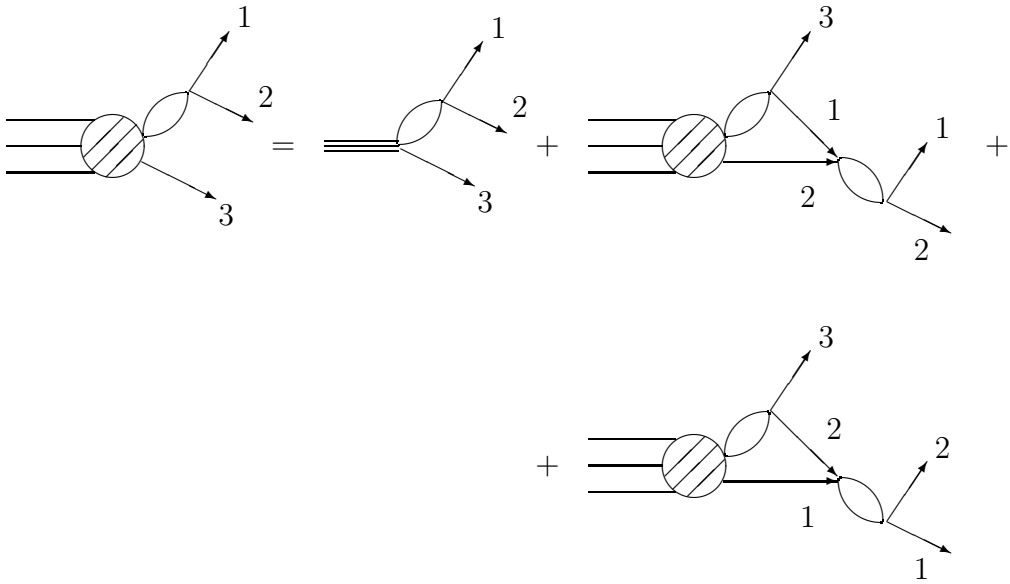
\put(-10,-10){{\large Fig.2. Graphic representation of the equations
for the amplitude $A_1(s,s_{ik})$.}}
\end{picture}

\newpage
\vskip60pt
\begin{picture}(600,80)
\multiput(70,20)(0,5){20}{\line(0,1){2}}
\put(-10,50){\line(1,0){33}}
\put(-10,60){\line(1,0){28}}
\put(-10,70){\line(1,0){33}}
\put(19,56){\line(1,1){15}}
\put(22,51){\line(1,1){17}}
\put(27.5,48.5){\line(1,1){14}}
\put(41,54){\vector(1,0){43}}
\put(30,60){\circle{25}}
\put(58,64){\oval(33,33)[tl]}
\put(42,80){\oval(33,33)[br]}
\put(59,81){\vector(2,3){15}}
\put(59,81){\vector(1,-1){25}}
\put(77,105){$3$}
\put(75,70){$1, 2$}
\put(60,37){$2, 1$}
\put(90,90){$k_{13}, k_{23}$}
\put(85,39){\oval(33,33)[tr]}
\put(101,55){\oval(33,33)[bl]}
\put(103,39){\vector(2,3){15}}
\put(103,39){\vector(2,-1){24}}
\put(121,40){$k_{12}\frac{A_1^0+3A_0^1}{4}\left|_{k_{12}}\right.$}
\put(200,70){$+$}
\multiput(305,20)(0,5){20}{\line(0,1){2}}
\put(225,50){\line(1,0){33}}
\put(225,60){\line(1,0){28}}
\put(225,70){\line(1,0){33}}
\put(254,56){\line(1,1){15}}
\put(257,51){\line(1,1){17}}
\put(262.5,48.5){\line(1,1){14}}
\put(276,54){\vector(1,0){43}}
\put(265,60){\circle{25}}
\put(293,64){\oval(33,33)[tl]}
\put(277,80){\oval(33,33)[br]}
\put(294,81){\vector(2,3){15}}
\put(294,81){\vector(1,-1){25}}
\put(312,105){$2$}
\put(310,70){$1, 3$}
\put(295,37){$3, 1$}
\put(325,90){$k_{12}, k_{23}$}
\put(320,39){\oval(33,33)[tr]}
\put(336,55){\oval(33,33)[bl]}
\put(338,39){\vector(2,3){15}}
\put(338,39){\vector(2,-1){24}}
\put(356,40){$k_{13}\frac{A_1^{0S}+3A_0^{1S}}{4}\left|_{k_{13}}\right.$}
\put(430,70){$+$}
\end{picture}

\vskip60pt
\begin{picture}(600,80)
\multiput(170,20)(0,5){20}{\line(0,1){2}}
\put(65,70){$+$}
\put(90,50){\line(1,0){33}}
\put(90,60){\line(1,0){28}}
\put(90,70){\line(1,0){33}}
\put(119,56){\line(1,1){15}}
\put(122,51){\line(1,1){17}}
\put(127.5,48.5){\line(1,1){14}}
\put(141,54){\vector(1,0){43}}
\put(130,60){\circle{25}}
\put(158,64){\oval(33,33)[tl]}
\put(142,80){\oval(33,33)[br]}
\put(159,81){\vector(2,3){15}}
\put(159,81){\vector(1,-1){25}}
\put(177,105){$1$}
\put(175,70){$2, 3$}
\put(160,37){$3, 2$}
\put(190,90){$k_{12}, k_{13}$}
\put(185,39){\oval(33,33)[tr]}
\put(201,55){\oval(33,33)[bl]}
\put(203,39){\vector(2,3){15}}
\put(203,39){\vector(2,-1){24}}
\put(221,40){$k_{23}A_1^{0S}\left|_{k_{23}}\right.$}
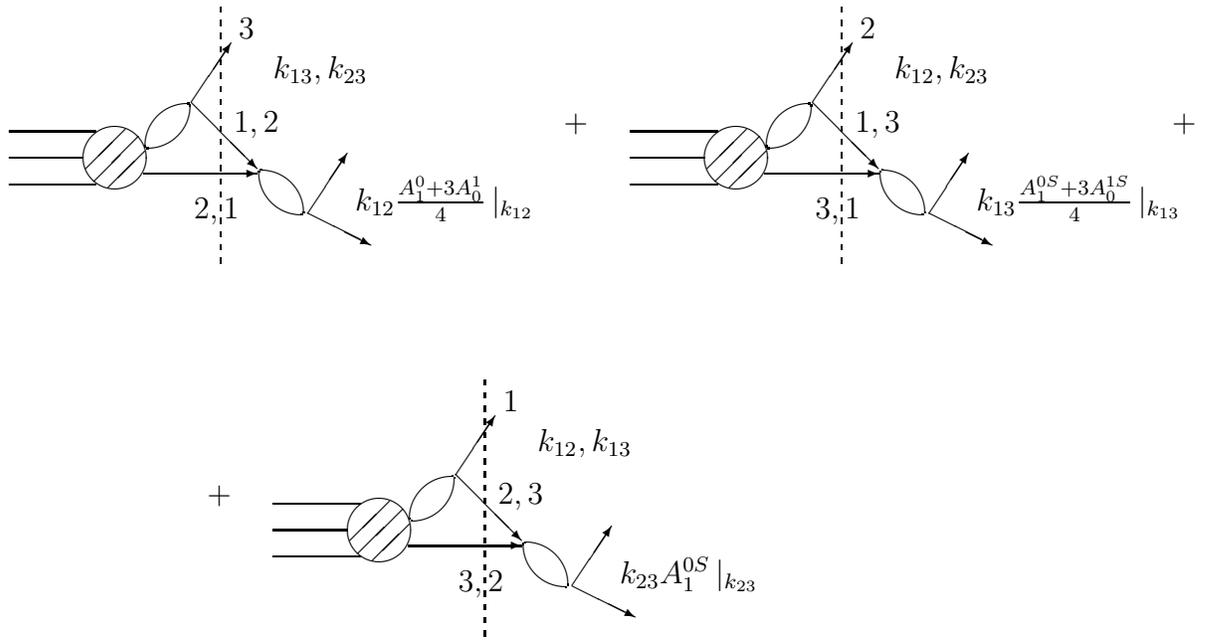
\put(-10,0){{\large Fig.3. The contribution of the diagrams with the
rescattering.}}
\end{picture}

\newpage
\noindent
{\large Table I.}

\noindent
{The $\Delta$-isobar masses of multiplet $(70,1^-)$.}

\vskip1.5ex

\noindent
\begin{tabular}{|c|c|c|c|}
\hline
Multiplet & Baryon & Mass ($GeV$) & Mass ($GeV$) (exp.) \\
\hline
$\frac{3}{2}^-$ $(10,2)$ & $D_{33}$ & $1.565$ & $1.700$\\
\hline
$\frac{1}{2}^-$ $(10,2)$ & $S_{31}$ & $1.650$ & $1.620$\\
\hline
\end{tabular}

\vskip1.5ex

\noindent
{The parameters of model (Tables I-VI): gluon coupling constant $g_+ =0.69$,
\\
$g_- =0.3$, cutoff energy parameters $\lambda=14.5$, $\lambda_s=11.2$.}

\vskip30pt

\noindent
{\large Table II.}

\noindent
{The nucleon masses of multiplet $(70,1^-)$.}

\vskip1.5ex

\noindent
\begin{tabular}{|c|c|c|c|}
\hline
Multiplet & Baryon & Mass ($GeV$) & Mass ($GeV$) (exp.) \\
\hline
$\frac{5}{2}^-$ $(8,4)$ & $D_{15}$ & $1.606$ & $1.675$\\
\hline
$\frac{3}{2}^-$ $(8,4)$ & $D_{13}$ & $1.565$ & $1.700$\\
\hline
$\frac{1}{2}^-$ $(8,4)$ & $S_{11}$ & $1.650$ & $1.650$\\
\hline
$\frac{3}{2}^-$ $(8,2)$ & $D_{13}$ & $1.520$ & $1.520$\\
\hline
$\frac{1}{2}^-$ $(8,2)$ & $S_{11}$ & $1.535$ & $1.535$\\
\hline
\end{tabular}

\vskip30pt

\noindent
{\large Table III.}

\noindent
{The $\Sigma$-hyperon masses of multiplet $(70,1^-)$.}

\vskip1.5ex

\noindent
\begin{tabular}{|c|c|c|c|}
\hline
Multiplet & Baryon & Mass ($GeV$) & Mass ($GeV$) (exp.) \\
\hline
$\frac{3}{2}^-$ $(10,2)$ & $D_{33}$ & $1.675$ & --\\
\hline
$\frac{1}{2}^-$ $(10,2)$ & $S_{31}$ & $1.785$ & --\\
\hline
$\frac{5}{2}^-$ $(8,4)$ & $D_{15}$ & $1.726$ & $1.775$\\
\hline
$\frac{3}{2}^-$ $(8,4)$ & $D_{13}$ & $1.675$ & $1.670$\\
\hline
$\frac{1}{2}^-$ $(8,4)$ & $S_{11}$ & $1.785$ & $1.750$\\
\hline
$\frac{3}{2}^-$ $(8,2)$ & $D_{13}$ & $1.590$ & $1.580$\\
\hline
$\frac{1}{2}^-$ $(8,2)$ & $S_{11}$ & $1.639$ & $1.620$\\
\hline
\end{tabular}

\vskip30pt

\noindent
{\large Table IV.}

\noindent
{The $\Lambda$-hyperon masses of multiplet $(70,1^-)$.}

\vskip1.5ex

\noindent
\begin{tabular}{|c|c|c|c|}
\hline
Multiplet & Baryon & Mass ($GeV$) & Mass ($GeV$) (exp.) \\
\hline
$\frac{5}{2}^-$ $(8,4)$ & $D_{15}$ & $1.634$ & --\\
\hline
$\frac{3}{2}^-$ $(8,4)$ & $D_{13}$ & $1.581$ & --\\
\hline
$\frac{1}{2}^-$ $(8,4)$ & $S_{11}$ & $1.702$ & --\\
\hline
$\frac{3}{2}^-$ $(8,2)$ & $D_{13}$ & $1.568$ & --\\
\hline
$\frac{1}{2}^-$ $(8,2)$ & $S_{11}$ & $1.617$ & --\\
\hline
$\frac{3}{2}^-$ $(1,2)$ & $D_{03}$ & $1.520$ & $1.520$\\
\hline
$\frac{1}{2}^-$ $(1,2)$ & $S_{01}$ & $1.437$ & $1.405$\\
\hline
\end{tabular}

\vskip40pt

\noindent
{\large Table V.}

\noindent
{The $\Xi$-hyperon masses of multiplet $(70,1^-)$.}

\vskip1.5ex

\noindent
\begin{tabular}{|c|c|c|c|}
\hline
Multiplet & Baryon & Mass ($GeV$) & Mass ($GeV$) (exp.) \\
\hline
$\frac{3}{2}^-$ $(10,2)$ & $D_{33}$ & $1.765$ & --\\
\hline
$\frac{1}{2}^-$ $(10,2)$ & $S_{31}$ & $1.905$ & --\\
\hline
$\frac{5}{2}^-$ $(8,4)$ & $D_{15}$ & $1.823$ & --\\
\hline
$\frac{3}{2}^-$ $(8,4)$ & $D_{13}$ & $1.765$ & $1.820$\\
\hline
$\frac{1}{2}^-$ $(8,4)$ & $S_{11}$ & $1.905$ & --\\
\hline
$\frac{3}{2}^-$ $(8,2)$ & $D_{13}$ & $1.681$ & --\\
\hline
$\frac{1}{2}^-$ $(8,2)$ & $S_{11}$ & $1.741$ & --\\
\hline
\end{tabular}

\vskip40pt

\noindent
{\large Table VI.}

\noindent
{The $\Omega$-hyperon masses of multiplet $(70,1^-)$.}

\vskip1.5ex

\noindent
\begin{tabular}{|c|c|c|c|}
\hline
Multiplet & Baryon & Mass ($GeV$) & Mass ($GeV$) (exp.) \\
\hline
$\frac{3}{2}^-$ $(10,2)$ & $D_{33}$ & $1.889$ & --\\
\hline
$\frac{1}{2}^-$ $(10,2)$ & $S_{31}$ & $2.040$ & --\\
\hline
\end{tabular}

\vskip50pt

\noindent
{\large Table VII. Coefficient of Ghew-Mandelstam function for the
different diquarks.}

\vskip1.5ex

\begin{tabular}{|c|c|c|c|}
\hline
 &$\alpha_J$&$\beta_J$&$\delta_J$\\
\hline
 & & & \\
$1^+$&$\frac{1}{3}$&$\frac{4m_i m_k}{3(m_i+m_k)^2}-\frac{1}{6}$
&$-\frac{1}{6}(m_i-m_k)^2$\\
 & & & \\
$0^+$&$\frac{1}{2}$&$-\frac{1}{2}\frac{(m_i-m_k)^2}{(m_i+m_k)^2}$&0\\
 & & & \\
$0^-$&$0$&$\frac{1}{2}$&$-\frac{1}{2}(m_i-m_k)^2$\\
 & & & \\
$1^-$&$\frac{1}{2}$&$-\frac{1}{2}\frac{(m_i-m_k)^2}{(m_i+m_k)^2}$&0\\
 & & & \\
$2^-$&$\frac{3}{10}$&$\frac{1}{5}
\left(1-\frac{3}{2}\frac{(m_i-m_k)^2}{(m_i+m_k)^2}\right)$
&$-\frac{1}{5}(m_i-m_k)^2$\\
 & & & \\
\hline
\end{tabular}

}
\end{document}